\documentclass{article}[12pt] 
\usepackage{ amsmath, amsthm, amssymb } 
\usepackage{axodraw} 
\usepackage{color} 
\usepackage{cite}

\newcommand{\be} { \begin{equation} } 
\newcommand{\ee} { \end{equation} } 

\newcommand{\labbel}[1] { \label{#1} } 

\newcommand{\Cla}{{\rm Cl}_2\left( \frac{\pi}{3}\right)}
\newcommand{\Lst}{{\rm Ls}_3\left( \frac{2\,\pi}{3}\right)}
\newcommand{\bb}{ (\sqrt{u}-1)^2 }
\newcommand{\BB}{ (\sqrt{u}+1)^2 }

\newcommand{\noo}{{(9,\infty)}}
\newcommand{\zu}{{(0,1)}}
\newcommand{\un}{{(1,9)}}
\newcommand{\ooz}{{(-\infty,0)}}

\newcommand{\Tad}{{\rm Tad}} 
\newcommand{\Bub}{{\rm Bub}} 
\newcommand{\Tri}{{\rm Tri}} 

\newcommand{\D}{\mathfrak{D}}
\newcommand{\I}{\mathcal{I}}

\newcommand{\Ima}{{\rm Im}}
\newcommand{\Rea}{{\rm Re}}
\newcommand{\Li}{{\rm Li}}
\newcommand{\intk} { \int {\mathfrak D}^d k } 
\newcommand{\intl} { \int {\mathfrak D}^d l } 

\newcommand{\eps} {\epsilon} 

\newcommand{\bubble}[1]{ 
\mbox{\parbox{2.5cm}{\hspace{0.25cm} 
\begin{picture}(2,1) 
\thicklines 
\put(0.3,0.5){\vector(1,0){0.1}} 
\put(-0.03,0.5){\line(1,0){0.5}} 
\put(1.53,0.5){\line(1,0){0.5}} 
\put(1,0.5){\circle{2}} 
\put(0.85,1.12){$m_1$} 
\put(0.85,0.08){$m_2$} 
\put(0.25,0.7){\makebox(0,0)[b]{$#1$}} 
\end{picture} 
}} 
\hfill}

\newcommand{\triangleo}[3]{
\mbox{\parbox{3cm}{\hspace{-0.1cm}
\begin{picture}(2.5,1.4)
\thicklines
\put(0,0.7){\line(1,0){0.5}}
\put(1.0,1.2){\line(1,0){0.7}}
\put(1.0,0.2){\line(1,0){0.7}}
\put(1.0,0.2){\line(0,1){1.0}}
\put(0.3,0.7){\vector(1,0){0.1}}
\put(1.5,0.2){\vector(1,0){0.1}}
\put(1.5,1.2){\vector(1,0){0.1}}
\put(0.5,0.7){\line(1,1){0.5}}
\put(0.5,0.7){\line(1,-1){0.5}}
\put(1,1.2){\line(1,0){0.5}}
\put(1,0.2){\line(1,0){0.5}}
\thinlines
\put(0.25,0.9){\makebox(0,0)[b]{$#1$}}
\put(1.85,1.2){\makebox(0,0)[l]{$#2$}}
\put(1.85,0.2){\makebox(0,0)[l]{$#3$}}
\end{picture}
}}
\hfill}

\newcommand{\sunrisetwo}[1]{ 
\mbox{\parbox{2.5cm}{\hspace{0.25cm} 
\begin{picture}(2,1) 
\thicklines 
\put(0.3,0.5){\vector(1,0){0.1}} 
\put(0.5,0.5){\line(1,0){1}} 
\put(0,0.5){\line(1,0){0.5}} 
\put(1.5,0.5){\line(1,0){0.5}} 
\put(1,0.5){\circle{2}} 
\put(0.85,1.12){$m_1$} 
\put(0.85,0.60){$m_2$} 
\put(0.85,0.08){$m_3$} 
\put(0.25,0.7){\makebox(0,0)[b]{$#1$}} 
\end{picture} 
}} 
\hfill}

\newcommand{\kite}[1]{ 
\mbox{\parbox{2.5cm}{\hspace{0.25cm} 
\begin{picture}(2,1) 
\thicklines
\put(0.3,0.5){\vector(1,0){0.1}} 
\put(0,0.5){\line(1,0){0.5}} 
\DashLine(18,18)(36,0){4}
\DashLine(37,37)(55,18){4}
\put(1.0,0.0){\line(0,1){1}} 
\put(1.5,0.5){\line(1,0){0.5}} 
\put(0.5,0.5){\line(1,1){0.5}} 
\put(1.0,0.0){\line(1,1){0.5}} 
\put(0.25,0.7){\makebox(0,0)[b]{$#1$}} 
\end{picture} 
}} 
\hfill} 

\begin{document} 
\setlength{\unitlength}{1.3cm} 
\begin{titlepage}
\vspace*{-1cm}
\begin{flushright}
TTP16-005
\end{flushright}                                
\vskip 3.5cm
\begin{center}
\boldmath
 
{\Large\bf Differential equations and dispersion relations for Feynman \\ 
amplitudes.
The two-loop massive sunrise and the kite integral\\[3mm] }
\unboldmath
\vskip 1.cm
{\large Ettore Remiddi}$^{a,}$
\footnote{{\tt e-mail: ettore.remiddi@bo.infn.it}}
{\large and Lorenzo Tancredi}$^{b,}$
\footnote{{\tt e-mail: lorenzo.tancredi@kit.edu}} 
\vskip .7cm
{\it $^a$ DIFA, Universit\`a di Bologna and INFN, Sezione di Bologna, 
I-40126 Bologna, Italy } \\
{\it $^b$ Institute for Theoretical Particle Physics, KIT, 76128 Karlsruhe, 
Germany } 
\end{center}
\vskip 2.6cm

\begin{abstract}
It is shown that the study of the imaginary part and of the corresponding 
dispersion relations of Feynman graph amplitudes 
within the differential equations method can provide a powerful tool for the 
solution of the equations, especially in the massive case. 
\par 
The main features of the approach are illustrated by discussing the 
simple cases of the 1-loop self-mass and of a particular vertex amplitude, 
and then used for the evaluation of the two-loop massive sunrise and the 
QED kite graph (the problem studied by Sabry in 1962), up to first order 
in the $(d-4)$ expansion. 

\end{abstract}
\vfill
\end{titlepage}                                                                
\newpage

\section{Introduction} \labbel{sec:intro} \setcounter{equation}{0} 
\numberwithin{equation}{section}

In the last years we have assisted to an impressive increase in our 
knowledge of the mathematical
structures that appear in multiloop Feynman integrals, thanks to the 
combined use of various computational techniques, such as 
to the method of differential 
equations~\cite{Kotikov:1990kg,Remiddi:1997ny,Gehrmann:1999as}, 
the introduction of a class of special functions, (dubbed originally harmonic 
polylogarithms, HPLs~\cite{Remiddi:1999ew,Gehrmann:2000zt}, 
they came out to be a subset of the much larger class of 
multiple polylogarithms, MPLs, 
see~\cite{Goncharov:1998kja,Goncharov:2001iea,Ablinger:2013cf,Panzer:2014caa} 
and references therein), the definition of a so-called canonical 
basis~\cite{Henn:2013pwa} for dealing with increasingly larger systems 
of differential equations and the use of the Magnus exponentiation~\cite{Argeri:2014qva}.\par 
However, most of the above results have been obtained in the massless 
limit; indeed, the situation for massive amplitudes is different, 
as the two-loop massive sunrise (which has three propagators only) 
is still the object of thorough 
investigation~\cite{Caffo:1998du,Laporta:2004rb,Bloch:2013tra,Remiddi:2013joa,
Adams:2013nia,Adams:2014vja,Adams:2015gva,Adams:2015ydq,Bloch:2016izu}. 
A general approach to the study of arbitrarily complicated systems of differential equations
within difference field theory has been recently proposed in~\cite{Ablinger:2015tua}.
\par 
In this paper we will show that the study of the imaginary parts 
and related dispersion relations satisfied by the Feynman amplitudes, 
within the differential equation frame, can provide another useful 
practical tool for their evaluation in the massive case as well. 
\par 
The imaginary parts of Feynman graphs can be obtained in various ways. 
To start with, one can use Cutkosky-Veltman 
rule~\cite{Cutkosky:1960sp,Veltman:1963th,Remiddi:1981hn} 
for integrating directly the loop momenta in the very definition of the 
graphs. When the $d$-continuous dimensional regularization is 
used, nevertheless, that is practical only in the simplest cases. Another possibility is 
the extraction of the imaginary part from the solution of the differential 
equations, which of course requires the knowledge of the solution itself. 
More interestingly, one can observe that often the differential equations become 
substantially simpler when restricted to the imaginary part only, so that their 
solution can become easier. \par 
In any case, once the imaginary part of some 
amplitude $ A(d;u) $, say $ {\Ima}A(d;u) $,  is obtained, one has at disposal the dispersive 
representation for $ A(d;u) $, namely an expression of the form 
$$ A(d;u) = \frac{1}{\pi}\int dt\ {\Ima}A(d;t)\ \frac{1}{t-u} $$ 
(where the limit of integration have been skipped for ease of typing). 
Such a representation turns out to be very useful when the amplitude 
$ A(d;u) $ appears within the inhomogeneous terms of some other differential 
equation, regardless of the actual analytical expression of $ A(d;u) $. 
Indeed, as the whole dependence on $ u $ is in the denominator $(t-u)$ 
one can work out its contribution by considering only that denominator, 
freezing, so to say,
the $ t $-integration and the 
weight $ {\Ima}A(d;t) $ until the dependence on the variable $ u $ 
(the variable of the differential equation) 
has been properly processed. Let us emphasize, again, that such a processing 
is, obviously, fully independent of the actual form of $ {\Ima}A(d;t) $. 
\par 
In the following, we will illustrate the above remarks in a couple of 
elementary applications and then use them in the case of the two-loop 
QED-kite, {\it i.e.} the two-loop electron self-mass in QED, already 
studied by Sabry~\cite{Sabry} long ago.
 The study of the kite amplitudes requires
in turn the knowledge of the two-loop massive sunrise, which appears as inhomogeneous 
terms in their differential equations.
Indeed, the imaginary part~\cite{Broadhurst:1987ei} and related 
dispersion relations~\cite{Bauberger:1994by,Bauberger:1994hx} have been already 
exploited long ago for studying the zeroth order of the sunrise 
and the kite integral. In this paper our goal is more general, as we will show how to use them 
consistently within the  differential equations approach, which will allow us to investigate 
the solution at any order in the $(d-4)$ expansion.
\par 

The paper is organized as follows.
We begin in section~\ref{sec:1lB} studying the imaginary part of the one-loop self mass
and its dispersion relation for generic values of the dimensions $d$. We elaborate
on its calculation both from Cutkosky-Veltman rule and from the differential equations.
In section~\ref{sec:1lT} we study a particular vertex amplitude
through the differential equations method. The one-loop self-mass appears as inhomogenous
term in the equations and we show that their evaluation can be simplified, 
once the one-loop self-mass is inserted as dispersive relation.
In section~\ref{sec:iter} we move our attention to the two-loop sunrise graph, which
we write as iteration of two one-loop bubbles. This allows us to derive an extremely 
compact representation valid for generic $d$, from which one can show that, 
at every order in $(d-2)$ (and therefore also in $(d-4)$), 
the sunrise can be written as a one-dimensional integral over a square root of a quartic polynomial, 
times a combination of multiple polylogarithms only. 
The simplicity of this result motivates us to look more systematically for a similarly simple 
structure using differential equations
from the very beginning for the whole integral family of the kite. In section~\ref{sec:deq}
we discuss the notation and describe the master integrals which have to be computed. 
In section~\ref{sec:simple} we provide the solution of the simple topologies, which can be written
in terms of HPLs only. Then in section~\ref{sec:basis} we start a systematic study of the differential equations
of the sunrise graph. It is known that the solution for the sunrise 
graph is somewhat simpler when its Laurent series is considered in $(d-2)$ 
instead of in $(d-4)$; however, we find more convenient to expand all the 
master integrals in $(d-4)$ from the very beginning. 
To that aim, by using the well known fact that any Feynman integral 
in $d-2$ dimensions can be written as a linear combination of integrals in 
$d$ dimensions, we build up a new, equivalent basis of master integrals 
for the sunrise whose expansion in $(d-4)$ is identical to the expansion 
of the original masters in $(d-2)$.
Once we have a convenient basis and the corresponding differential
equations, we show how to solve them iteratively in section~\ref{sec:sun}. 
We conclude the section 
providing explicit analytical results for both master integrals for the first two non-zero orders 
and showing
how to extract their imaginary parts and write dispersion relations for them.
We move then to the kite integral in section~\ref{sec:kite}, where we show how
the representation of the sunrise as a dispersion relation is particularly convenient, as it allows 
to write a compact  solution
for the first two orders of the kite integral.
Finally we conclude in section~\ref{sec:concl}. We enclose different appendices
where we provide further mathematical details and explicit derivations.

\section{The 1-loop self-mass: imaginary part and dispersion relation } 
\labbel{sec:1lB} \setcounter{equation}{0} 
\numberwithin{equation}{section} 
We define the integration over a loop momentum $k$ in $d$ 
continuous dimensions as 
\be \intk = \frac{1}{C(d)} \int \frac{d^d k}{(2 \pi)^{d-2}}\,, 
\labbel{defDk} \ee 
with 
\be C(d) = (4 \pi)^{(4-d)/2} \Gamma\left( 3 - \frac{d}{2}\right) \ , 
\labbel{defCd} \ee 
so that the tadpole amplitude $ \Tad(d;m) $ reads 
\be \Tad(d;m) = \intk \ \frac{1}{k^2+m^2} 
           =  \frac{m^{d-2}}{(d-2)(d-4)} \ . \labbel{defTad} \ee 
We then consider the 1-loop ``bubble" 
\begin{align} 
 \Bub(d;-q^2,m_1,m_2) &= \bubble{q} \nonumber\\ 
          &= \intk \ \frac{1}{(k^2+m_1^2)((q-k)^2+m_2^2)} \,. 
\labbel{defB} \end{align} 
We work in the Euclidean metric such that $q^2$ is positive when $q$ is spacelike. 
At $ q = 0 $ one has at once 
\be \Bub(d;0,m_1,m_2) = \frac{1}{m_1^2-m_2^2} 
          \left( \Tad(d;m_2) - \Tad(d;m_1) \right) \ . \labbel{B(d,0)} \ee
\par 
Cutkosky-Veltman rule gives for the imaginary part of the bubble 
amplitude in $d$-continuous dimensions and for $ s = -q^2 > (m_1+m_2)^2 $ 
the expression 
\begin{align}
  {\Ima}\Bub(d;s,m_1,m_2) = \pi \ \frac{1}{2} 
  \frac{B_d}{\sqrt{R_2(s,m_1^2,m_2^2)}} 
    \,\left( \frac{R_2(s,m_1^2,m_2^2)}{s} \right)^{(d-2)/2} \ , 
                                          \labbel{ImBub} 
\end{align} 
where we introduced the usual K\"{a}llen function 
\begin{align} 
R_2(s,m_1^2,m_2^2) &= s^2+m_1^4+m_2^4-2m_1^2s-2m_2^2s-2m_1^2m_2^2 
\nonumber\\        &= (s-(m_1+m_2)^2)(s-(m_1-m_2)^2) \ , 
\labbel{defR2} \end{align} 
and the $d$ dependent coefficient 
\begin{align} 
B_d = \frac{4\sqrt\pi }{2^d\ \Gamma\left( 3 - \frac{d}{2}\right) 
\Gamma \left( \frac{d-1}{2}\right)}\,, \labbel{defBd} 
\end{align} 
whose expansion for $d \approx 2$ reads 
\begin{align} 
 B_d = 1 + \frac{1}{2} (d-2) - \frac{1}{12} 
       (\pi^2-3)(d-2)^2 + \mathcal{O}((d-2)^3)\,. \labbel{Bdat2} 
\end{align}
As a further remark, Eq.(\ref{ImBub}) can also be written as 
\be \frac{1}{\pi}\, {\Ima}B(d;s,m_1^2,m_2^2) = 
 \frac{B_d}{2} \left[s-(m_1+m_2)^2\right]^{\frac{d-3}{2}} 
               \left[s-(m_1-m_2)^2\right]^{\frac{d-3}{2}} 
               s^{-\frac{d-2}{2}} \ . \labbel{eq:ImB1} \ee 
Once the imaginary part is given, we can write a dispersion relation for the 
one-loop bubble
\begin{align} 
 \Bub(d;-q^2,m_1,m_2) &= \intk 
                   \ \frac{1}{(k^2+m_1^2)((q-k)^2+m_2^2)} \nonumber\\ 
    &= \int_{(m_1+m_2)^2}^\infty \frac{dt}{t+q^2} 
                   \frac{1}{\pi}\, {\Ima}\Bub(d;s,m_1,m_2) \ . 
\labbel{eq:dispB} 
\end{align} 
Note that Eq.(\ref{ImBub}) is valid for arbitrary values 
of $ d $, but Eq.(\ref{ImBub}) is written in the form most convenient 
for the expansion in $ (d-2) $; the same holds also for 
Eq.(\ref{eq:dispB}), which is however convergent only for $ d<4 $. 
To obtain a formula valid also in the $ d\approx4 $ region, 
one can write a subtracted dispersion relation 
\begin{align}
 \Bub(d;-q^2,m_1,m_2) &= \Bub(d;0,m_1,m_2) 
 - q^2\int_{(m_1+m_2)^2}^\infty \frac{dt}{t(t+q^2)} 
              \frac{1}{\pi}\ {\Ima}\Bub(d;t,m_1^2,m_2^2) \ , 
\labbel{eq:subdispB} 
\end{align} 
where $ \Bub(d;0,m_1,m_2) $, which is given in Eq,(\ref{B(d,0)}), contains 
a pole at $ d=4 $, while the integral is convergent for $ d < 6 $. 
\par 
The 1-loop self-mass amplitude Eq.(\ref{defB}), which in the following 
will be written as $ \Bub(d;s) $ for ease of typing, is known to satisfy 
the following differential equation in $ s $ 
\begin{align} \frac{d}{ds} \Bub(d;s) = 
   &- \frac{1}{2}\left( \frac{1}{s-(m_1+m_2)^2} + \frac{1}{s-(m_1-m_2)^2} 
                 \right) \Bub(d;s) \nonumber\\ 
   &- \frac{1}{2}\left( \frac{1}{s} - \frac{1}{s-(m_1+m_2)^2} 
                 - \frac{1}{s-(m_1-m_2)^2} \right) (d-2)\Bub(d;s) \nonumber\\ 
   &+ N(d;s) \ , \labbel{EqBd} \end{align} 
where the inhomogeneous term, $ N(d;s) $ is given by 
\begin{align} N(d;s) &= \frac{d-2}{4m_1(m_1^2-m_2^2)} \left( 
      - 2\frac{m_1}{s} + \frac{m_1-m_2}{s-(m_1+m_2)^2} 
                       + \frac{m_1+m_2}{s-(m_1-m_2)^2} 
                      \right) \Tad(d;m_1) \nonumber\\ 
  &+ \frac{d-2}{4m_2(m_1^2-m_2^2)} \left( 
      + 2\frac{m_2}{s} + \frac{m_1-m_2}{s-(m_1+m_2)^2} 
                       - \frac{m_1+m_2}{s-(m_1-m_2)^2} 
                      \right) \Tad(d;m_2) \ . \labbel{defN} 
\end{align} 
The homogeneous equation associated to Eq.(\ref{EqBd}) is (obviously) 
\begin{align} \frac{d}{ds} b(d;s) = 
   &- \frac{1}{2}\left( \frac{1}{s-(m_1+m_2)^2} + \frac{1}{s-(m_1-m_2)^2} 
                 \right) b(d;s) \nonumber\\ 
   &- \frac{1}{2}\left( \frac{1}{s} - \frac{1}{s-(m_1+m_2)^2} 
                 - \frac{1}{s-(m_1-m_2)^2} \right) (d-2)b(d;s) \ . 
\labbel{Eqbd} \end{align} 
One sees immediately that $ {\Ima}\Bub(d;s) $, Eq.(\ref{eq:ImB1}) satisfies 
the homogeneous equation, for any value of $ d $. That fact is hardly 
surprising, yet it deserves some comments. \par 
When looking for a solution of an equation like Eq.(\ref{EqBd}), it can be 
convenient, in order to fix the boundary conditions, to start by considering 
values of the variable $ s $ for which the solution is expected to be 
real (typically, $ s=0 $ or $s $ negative, {\it i.e.} in the spacelike 
region). 
But as one is also interested in the value of the solution for timelike, 
physical values of $s $, one is naturally lead to consider the 
solution as a complex analitical function of the argument $s$, to 
be evaluated along the whole line $ s+i\epsilon $, with 
$ s $ real and varying in the range $ -\infty < s < +\infty $ 
and $ \epsilon $ small and positive (the Feynman prescription). 
As the singular points of the equation correspond to real values of 
$ s $, such as for instance $ s =(m_1\pm m_2)^2 $, the function has 
no singularities along the $ s+i\epsilon $ line, so that its 
value is fully determined by the analytic continuation in terms of the 
initial boundary conditions. \par 
Moreover, one might be interested in considering separately the real part 
$ {\Rea}\Bub(d;s) $ and the imaginary part $ {\Ima}\Bub(d;s) $ of the solution 
$ \Bub(d;s) = {\Rea}\Bub(d;s) + i{\Ima}\Bub(d;s) $. In so doing, as the 
inhomogeneous term is real, Eq.(\ref{EqBd}) splits into the two equations 
\begin{align} \frac{d}{ds} {\Rea}\Bub(d;s) =
   &- \frac{1}{2}\left( \frac{1}{s-(m_1+m_2)^2} + \frac{1}{s-(m_1-m_2)^2}
                 \right) {\Rea}\Bub(d;s) \nonumber\\
   &- \frac{1}{2}\left( \frac{1}{s} - \frac{1}{s-(m_1+m_2)^2}
          - \frac{1}{s-(m_1-m_2)^2} \right) (d-2){\Rea}\Bub(d;s) \nonumber\\
   &+ N(d;s) \labbel{EqReBd} \\ 
   \frac{d}{ds} {\Ima}\Bub(d;s) = 
   &- \frac{1}{2}\left( \frac{1}{s-(m_1+m_2)^2} + \frac{1}{s-(m_1-m_2)^2}
                 \right) {\Ima}\Bub(d;s) \nonumber\\
   &- \frac{1}{2}\left( \frac{1}{s} - \frac{1}{s-(m_1+m_2)^2}
             - \frac{1}{s-(m_1-m_2)^2} \right) (d-2){\Ima}\Bub(d;s) \ , 
\labbel{EqImBd} \end{align} 
where Eq.(\ref{EqImBd}) is of course identical to Eq.(\ref{Eqbd}). 
One can now look at the {\it real} solution of Eq.(\ref{EqImBd}) for 
{\it real} values of $ s $. One finds easily: \\ 
if $ 0 < s < (m_1-m_2)^2 $ the solution is 
\be {\Ima}B(d,s) = c_1 \left[(m_1+m_2)^2-s\right]^{\frac{d-3}{2}} 
               \left[(m_1-m_2)^2-s\right]^{\frac{d-3}{2}} 
               s^{-\frac{d-2}{2}} \ ; \nonumber \ee 
if $ (m_1-m_2)^2 < s < (m_1+m_2)^2 $ the solution is 
\be {\Ima}B(d,s) = c_2 \left[(m_1+m_2)^2-s\right]^{\frac{d-3}{2}} 
               \left[s-(m_1-m_2)^2\right]^{\frac{d-3}{2}} 
               s^{-\frac{d-2}{2}} \ ; \nonumber \ee 
if $ (m_1+m_2)^2 < s < +\infty $ the solution is 
\be {\Ima}B(d,s) = c_3 \left[s-(m_1+m_2)^2\right]^{\frac{d-3}{2}} 
               \left[s-(m_1-m_2)^2\right]^{\frac{d-3}{2}} 
               s^{-\frac{d-2}{2}} \ . \nonumber \ee 
The evaluation of the solutions in the various regions is almost trivial, 
but one needs the knowledge of three constants, $ c_1, c_2, c_3 $ to 
actually recover the imaginary part Eq.(\ref{eq:ImB1}) 
(in that case the constants are, 
obviously, $ c_1 = c_2 = 0 $, $ c_3 = \pi B_d/2 $). \par 
Summarising, the evaluation of the imaginary parts alone within the 
differential equation approach is much simpler than the evaluation of 
the complete solution (real and imaginary parts), but requires 
some additional external information (such as the knowledge of the 
regions in which the imaginary part vanishes and 
its normalization when not vanishing). 
\section{The 1-loop self-mass and the 1-loop equal mass triangle } 
\labbel{sec:1lT} \setcounter{equation}{0} 
\numberwithin{equation}{section} 
In this section we show how to use Eq.(\ref{eq:dispB}) in the solution 
of the differential equation for a particular massive triangle amplitude, 
namely 
\begin{align} 
\Tri(d;s) &= \triangleo{q}{p_1}{p_2}\nonumber \\ 
&= \int \D^d k \frac{1}{(k^2+m^2)((k-p_1)^2+m^2)((k-p_1-p_2)^2+m^2)} 
\labbel{defTri} \end{align} 
with $ q=p_1+p_2$, $\ p_1^2 = p_2^2 =0 $ and $ -q^2 = s$ ($q^2$ is positive 
when $ q $ is spacelike).
The differential equation in the variable $ s $ for the amplitude 
$ \Tri(d;s) $ reads
\begin{align}
\frac{d}{ds} \Tri(d;s) &= -\frac{1}{s} \Tri(d;s) +
\frac{(d-2)}{8 m^4}\left(  \frac{1}{s-4 m^2} - \frac{1}{s}\right)
\Tad(d;m)\nonumber \\
&+\frac{(d-3)}{4\,m^2} \left(  \frac{1}{s-4 m^2} - \frac{1}{s}\right)
                  \Bub(d;s) \ , 
\end{align} 
where $ \Tad(d;m) $ is the tadpole defined in Eq.(\ref{defTad}), 
and $ \Bub(d;s) $ is the equal mass limit of the 1-loop self-mass of the previous 
section, {\it i.e.} $ \Bub(d;s) = \Bub(d;s,m,m) $, see Eq.(\ref{defB}). 
\par 
Now we notice that the homogeneous part of the equation is independent of
$d$ and reads
$$\frac{d}{ds} h(s) = -\frac{1}{s} h(s) \ ; $$ 
its solution (apart from a multiplicative constant) is 
\begin{equation} 
 h(s) = \frac{1}{s}\,. \labbel{defh} 
\end{equation}
We can then use Euler's method to write the general solution for the triangle
as follows
\begin{align}
\Tri(d;s) = c(d;m) \frac{1}{s}\, 
&+ \frac{(d-2)}{2\,s} \int_0^s\,  \frac{du}{u-4 m^2} \Tad(d;m) \nonumber \\
&+ \frac{(d-3)}{s} \int_0^s\,  \frac{du}{u-4 m^2} \, \Bub(d;u) \,, 
\labbel{EulTri} \end{align} 
where $c(d,m)$ is an integration constant, depending in general 
on $d$ and $m$. 
We can fix the integration constant requiring that for $s \to 0 $ the 
amplitude is not divergent, which implies 
\be c(d;m) = 0\ . \ee 
The tadpole, Eq.(\ref{defTad}), is of course independent of $u$
and for the bubble we use its dispersive representation Eq.(\ref{eq:dispB})
\begin{align} 
\Bub(d;u) = \frac{1}{\pi} \int_{4 m^2}^\infty \frac{dt}{t-u - i\epsilon} 
 {\Ima}\Bub(d;t) \,. \labbel{dispBub} 
\end{align} 
Let us recall that the above integral is convergent for $ d\approx 2$ and that
if one is interested in $ d\approx 4 $, one can use its subtracted version 
Eq.(\ref{eq:subdispB}). 
Assuming for definiteness $ s < 4m^2 $, and therefore ignoring for the 
moment the $+i\epsilon$ prescription in Eq.(\ref{dispBub}), 
the triangle amplitude becomes 
\begin{align}
\Tri(d;s) &= \frac{m^{d-2}}{2\,s\,(d-4)}
\int_0^s\,  \frac{du}{u-4 m^2}  \nonumber \\
&+ \frac{(d-3)}{s} \frac{1}{\pi} \int_{4 m^2}^\infty \,\frac{dt}{t-4 m^2}\,
{\Ima}\Bub(d;t) 
\int_0^s\, du\, \left( \frac{1}{u-4 m^2} + \frac{1}{t-u}  \right)\,. 
\labbel{eq:Tri0} \end{align} 
The integration in $u$ is trivial and we get 
\begin{align} 
\Tri(d;s) &= \frac{m^{d-2}}{2\,s\,(d-4)}
\ln{\left( 1- \frac{s}{4 m^2}\right)}  \nonumber \\
&+ \frac{(d-3)}{s} \frac{1}{\pi} 
\int_{4 m^2}^\infty \, \frac{dt}{t-4 m^2}\, {\Ima}\Bub(d;t) 
\left[ \ln{\left( 1- \frac{s}{4 m^2}\right)} 
      -\ln{\left( 1- \frac{s}{t}\right)} \right] 
\,. \labbel{eq:Tri1l} 
\end{align} 
Note that the above result holds for any $d$ (within the considered range) 
independently of the actual 
explicit form of the inserted amplitude $ \Bub(d;u) $. 

If $0<s<4 m^2$ the result~\eqref{eq:Tri1l} is real, while for 
$s>4 m^2$ it develops an imaginary part. 
In order to properly extract it, it is enough to notice that, for $s>4 m^2$,
the $s\to s+ i\epsilon$ prescription gives 
\begin{align} 
\ln{\left( 1- \frac{s+i\epsilon}{4 m^2} \right)} = 
             \ln{\left( \frac{s}{4 m^2} -1 \right)} - i\, \pi\,,
\end{align} 
and 
\begin{align} 
\int_{4 m^2}^\infty \,  \frac{dt}{t-4 m^2}\, {\Ima}\Bub(d;t) 
           \ln{\left( 1- \frac{s+i\epsilon}{t}\right)} &= 
\int_{4 m^2}^s \,  \frac{dt}{t-4 m^2}\, {\Ima}\Bub(d;t) \,\left[ 
\ln{\left( \frac{s}{t} - 1 \right)} - i\, \pi\right]\nonumber \\ 
&+\int_s^\infty \, \frac{dt}{t-4 m^2}\, {\Ima}\Bub(d;t) 
\ \ln{\left( 1- \frac{s}{t}\right)} \ . 
\end{align} 
Collecting results and combining the various terms, the imaginary part 
of $ \Tri(d;s) $ for $ s>4m^2$ becomes 
\begin{align} 
\frac{1}{\pi}{\Ima}\Tri(d;s) = - \frac{m^{d-2}}{2\,s\,(d-4)}  
  - \frac{(d-3)}{s} \frac{1}{\pi} \int_s^\infty \, 
    \frac{dt}{t-4m^2} \, {\Ima} \Bub(d;t) \,. \labbel{ImTri} 
\end{align} 
It is to be noted, again, that the above result has been obtained 
from Eq.(\ref{eq:Tri1l}) independently of the explicit analytic 
expression of $ {\Ima}\Bub(d;u) $. 
As a check, we can write the dispersion relation for the triangle 
amplitude in terms of its imaginary part (we take $ s<4m^2 $ for 
simplicity) 
\begin{align} 
 \Tri(d;s) = \frac{1}{\pi} \int_{4 m^2}^\infty \frac{du}{u-s} 
             {\Ima} \Tri(d;u) \,. \labbel{dispTri} 
\end{align} 
By exchanging the order of integrations according to 
$$ \int_{4 m^2}^\infty \,du \, \int_u^\infty \, dt\, 
 = \int_{4 m^2}^\infty \,dt \, \int_{4 m^2}^t \, du\,,$$ 
Eq.(\ref{eq:Tri1l}) is easily recovered. 
\par 
Summarising, the use of the dispersive representation of the inserted 
amplitude $ \Bub(d;u) $ in the Euler form of the solution of the 
differential equation for the triangle amplitude gives, almost at once, 
the explicit form of $ \Tri(d;s) $, Eq.(\ref{eq:Tri1l}) in terms of 
$ {\Ima}\Bub(d;u) $. The imaginary part $ {\Ima}\Tri(d;s) $ of the triangle amplitude 
Eq.(\ref{ImTri}) can also be written in terms of $ {\Ima}\Bub(d;u) $, 
without explicit reference to the analytic form of the latter. 
The resulting dispersion relation Eq.(\ref{dispTri}) can be useful 
if $ \Tri(d;u)$ appears within the inhomogeneous terms of the equations 
for the amplitudes of some other process (such as for instance the 
QED light-light graphs). 

\section{The sunrise as iteration of the bubble graph} 
\labbel{sec:iter} \setcounter{equation}{0} 
\numberwithin{equation}{section}
Let us consider the sunrise scalar amplitude defined as
\vspace{0.3cm}
\begin{align}
S(d;-p^2,m_1,m_2,m_3) &= \sunrisetwo{p} \nonumber \\& = \intk \intl 
\ \frac{1}{(k^2+m_1^2)(l^2+m_2^2)((p-k-l)^2+m_3^2)}\,, \labbel{eq:sunscal} 
\end{align}
where the integration measure is defined in Eq.(\ref{defDk}) and we work in 
the Euclidean metric for simplicity.
It is well known that the sunrise graph with different masses possesses 
four master integrals,
which reduce to two in the case of equal masses~\cite{Caffo:1998du}.
In this section we will not try to give a full solution for all the masters 
integrals, but instead we will limit ourselves to considering the scalar 
integral~\eqref{eq:sunscal} only and try to study its iterative structure 
in $(d-2)$, or equivalently in $(d-4)$. 
One possible way to do this is by noting that the sunrise integral can be 
written as
\begin{align}
S(d;-p^2,m_1,m_2,m_3) = \intk \frac{1}{k^2+m_1^2} 
\intl \frac{1}{(l^2+m_2^2)((p-k-l)^2+m_3^2)}\,, \labbel{sun1} 
\end{align}
and according to Eq.(\ref{defB}) the integral in the momentum $l$ is simply 
a one-loop bubble with masses $m_2$ and $m_3$ and momentum $q = (p-k) $, 
\begin{equation} 
\Bub(d;-q^2,m_2,m_3) =  \intl \frac{1}{(l^2+m_2^2)((q-l)^2+m_3^2)}\,.
\end{equation}
The dispersive representation Eq.(\ref{eq:dispB}) then gives 
\be \intl \frac{1}{(l^2+m_2^2)((q-l)^2+m_3^2)} = 
   \int_{(m_2+m_3)^2}^\infty \frac{dt}{t+q^2} 
   \frac{1}{\pi}\ {\Ima}\Bub(d;t,m_2,m_3) \ , \ee 
with $ {\Ima}\Bub(d;t,m_2,m_3) $ given by Eq.(\ref{ImBub}). 
As $ q=p-k$, Eq.(\ref{sun1}) becomes 
\begin{align}
S(d;-p^2,m_1,m_2,m_3) &= \int_{(m_2+m_3)^2}^\infty dt 
   \ \frac{1}{\pi}\ {\Ima}\Bub(d;t,m_2,m_3) \nonumber\\ 
   &\times \intk \frac{1}{(k^2+m_1^2)((p-k)^2+t)} \,.
\end{align} 
Now clearly the integral in $k$ can be seen again as a one-loop 
bubble amplitudes, this time with (squared) masses $m_1^2$ and $t$. 
Using again the formula~\eqref{eq:dispB} we get 
\begin{align} 
S(d;-p^2,m_1,m_2,m_3) &= \int_{(m_2+m_3)^2}^\infty dt 
       \ \frac{1}{\pi}\ {\Ima}\Bub(d;t,m_2,m_3) \nonumber\\ 
  &\times \int_{(\sqrt{t}+m_1)^2}^\infty \frac{dv}{v+p^2} 
   \ \frac{1}{\pi}\ {\Ima}\Bub(d;v,\sqrt{t},m_1) 
\,. \labbel{eq:dispsun1}
\end{align}
We can obtain an equivalent representation by further exchanging the 
integrations in the variables $ t $ and $ v $ 
$$\int_{(m_2+m_3)^2}^\infty   dt \,\int_{(\sqrt{t}+m_1)^2}^\infty \, dv = 
\int_{(m_1+m_2+m_3)^2}^\infty \, dv 
\int_{(m_2+m_3)^2}^{(\sqrt{v}-m_1)^2} \, dt $$ 
such that, by recalling Eq.(\ref{ImBub}), we are left with
\begin{align} 
S(d;-p^2,m_1,&m_2,m_3) = 
\frac{B_d^2}{4}  \int_{(m_1+m_2+m_3)^2}^\infty   \frac{dv}{v+p^2}\nonumber \\ 
&\times \int_{(m_2+m_3)^2}^{(\sqrt{v}-m_1)^2} 
\frac{dt}{\sqrt{R_2(t,m_2^2,m_3^2) R_2(v,t,m_1^2)}} 
 \left( \frac{R_2(t,m_2^2,m_3^2)}{t} \frac{R_2(v,t,m_1^2)}{v} 
  \right)^{(d-2)/2}\,. \labbel{eq:dispsun2} 
\end{align} 
Eq.(\ref{eq:dispsun2}) is the main result of this section. From it we 
obtain at once, when $ -p^2 = s > (m_1+m_2+m_3)^2, $ 
\begin{align}
\frac{1}{\pi}{\Ima}&S(d;s,m_1,m_2,m_3) \nonumber \\ &= 
\frac{B_d^2}{4} \int_{(m_2+m_3)^2}^{(\sqrt{s}-m_1)^2}
\frac{dt}{\sqrt{R_2(t,m_2^2,m_3^2) R_2(s,t,m_1^2)}}
\left( \frac{R_2(t,m_2^2,m_3^2)}{t} 
  \frac{R_2(s,t,m_1^2)}{s}\right)^{(d-2)/2}\,. 
\labbel{eq:impartsun}
\end{align}
Note that 
Eq.\eqref{eq:impartsun} is nothing but the 
$d$-dimensional three-body massive phase space and Eq.(\ref{eq:dispsun2}) 
could indeed have been obtained also by computing first 
the imaginary part of the sunrise graph using Cutkosky-Veltman rule, 
and then writing a dispersion  relation for it. 
Remarkably, the complexity of the result in the general mass case 
is practically the same as in the equal mass case $m_1=m_2=m_3=m$. 
\par 
Let us further emphasize that Eqs.~\eqref{eq:dispsun1},~\eqref{eq:dispsun2} 
and~\eqref{eq:impartsun} are all true for generic, continuous values of 
$d$. Furthermore, their expansion in $(d-n)$, where $n$ is virtually any 
positive integer (and in particular in $(d-2)$), is completely
straightforward and generates only products of logarithms\footnote{Note that 
in odd numbers of dimensions, $d=2\,n + 1$, the imaginary part becomes particularly simple
since the square root in Eq.\eqref{eq:impartsun} cancels.}. 
This implies in turn that, at every order in $(d-2)$, the integral
in $ v $ in Eq.\eqref{eq:dispsun1} can 
\textsl{always be performed} in terms of multiple polylogarithms only.
This shows that, at every order in $(d-2)$, the sunrise integral can be written
as a one-fold integral over the root of a quartic polynomial, times combinations
of multiple polylogaritms. The result is interesting and it resembles 
similar results found for the finite term of a completely unrelated 
massless double box in $\mathcal{N}=4$~\cite{Paulos:2012nu,
CaronHuot:2012ab}\footnote{One should compare in particular our Eq.\eqref{eq:dispsun1}
with Eq.(3.23) in~\cite{CaronHuot:2012ab}.}.
Finally, the relation of this representation of the imaginary part of the sunrise
Eq.\eqref{eq:impartsun} with the results 
obtained by the explicit solution of the system of differential equations 
for the two amplitudes of the sunrise problem (which involves two 
pairs of solutions, {\it i.e.} four functions altogether, 
see for instance section~\ref{sec:sun} of this paper) is also intriguing, but will not 
be further investigated here. Starting from the next section we will instead focus
on the more general problem of computing the full set of master integrals
of the kite graph using the differential equations method.

\section{The differential equations for the kite master integrals} 
\labbel{sec:deq} \setcounter{equation}{0} 
\numberwithin{equation}{section} 
Let us consider the family of the integrals of the QED kite graph 
with three massive propagators and two massless ones, defined as 
\begin{align}
\I(n_1,n_2,n_3,n_4,n_5) &= \kite{p} \nonumber \\ &= 
\int \D^d k \,\D^d l\, \frac{1}{D_1^{n_1} D_2^{n_2} D_3^{n_3} 
                                          D_4^{n_4} D_5^{n_5}}
\end{align}
where dashed lines represent massless propagators. The five denominators are chosen as
\begin{align}
&D_1 = k^2 + m^2\,, \qquad D_2 = l^2 \,, 
                    \qquad D_3 = (k-l)^2 + m^2\,,\nonumber \\
&D_4 = (k-p)^2\,, \qquad D_5 = (l-p)^2 + m^2\,, \label{topo}
\end{align}
with $ -p^2 = s $ and $ p^2>0 $ when $ p $ is spacelike. 
The integration measure is defined as in Eq.(\ref{defDk}) 
such that according to Eq.(\ref{defTad}) the one-loop tadpole reads 
\begin{align}
\int \frac{\D^d k}{k^2 + m^2} = \frac{m^{d-2}}{(d-2)(d-4)}\,.
\end{align}
The integral family~\eqref{topo} can be very easily reduced to master integrals
using, for example, Reduze 2~\cite{Studerus:2009ye,vonManteuffel:2012np}. 
In order to simplify the notation
we put $m=1$ and define $u = s/m^2 $.
We find $8$ independent master integrals which we choose as follows
\begin{align}
&M_1(d;u) = \I(2,0,2,0,0)\,, \quad M_2(d;u) = \I(2,0,2,1,0)\,, \nonumber \\
&M_3(d;u) = \I(0,2,2,1,0)\,, \quad M_4(d;u) = \I(0,2,1,2,0)\,, \nonumber \\
&M_5(d;u) = \I(2,1,0,1,2)\,, \quad M_6(d;u) = \I(1,0,1,0,1)\,, \nonumber \\
&M_7(d;u) = \I(2,0,1,0,1)\,, \quad M_8(d;u) = \I(1,1,1,1,1)\,.
\end{align}
Most of the master integrals are very simple and have been already studied 
thoroughly
in the literature. In particular $M_1$,...,$M_5$ are known and 
can be written in terms of HPLs only. 
The remaining three integrals, $M_6$, $M_7$ and $M_8$, cannot be expressed 
in terms of MPLs and will be the main topic of this paper.  Note that 
$M_6$ and $M_7$ are the two master integrals of the two-loop massive 
sunrise with equal masses, see Eq.\eqref{eq:sunscal}. As we will see, 
$ M_6 $ and $ M_7 $ satisfy a system of two coupled differential 
equations, with $ M_6 $ appearing further within the 
inhomogeneous terms of the differential equation for $ M_8 $. 
\par 
As usual, we are interested in the Laurent expansion of the master 
integrals for $d \approx 4$. The computation of the first five integrals 
in terms of HPLs is straightforward.
In particular, it can be simplified by the choice of a canonical basis 
in $d\approx4$, which can be found  following the methods described 
in~\cite{Gehrmann:2014bfa,Henn:2014qga}. 
For the last three integrals, instead, a canonical basis in the usual sense 
cannot be found and we will have to resort to different arguments in order 
to put the system of differential equations in a 
form that is suitable for their integration.
We choose the following canonical basis for the simple topologies
\begin{align}
&f_1(d;u) = 4\,(d-4)^2 \,M_1(d;u)\,, \quad
f_2(d;u) = (d-4)^2\, u\, M_2(d;u)\,, 
\nonumber \\
&f_3(d;u) = (d-4)^2\, u\, M_3(d;u)\,, \quad f_4(d;u) = (d-4)^2\, 
          (1-u)\left[ \frac{1}{2}\,M_3(d;u) +  M_4(d;u)\right]\,, \nonumber \\
&f_5(d;u) = (d-4)^2\, u^2\, M_5(d;u)\,, \labbel{eq:basisca}
\end{align}
while for the non-trivial topologies we introduce 
\begin{align}
&f_6(d;u) = (d-4)^2\, M_6(d;u)\,, \nonumber\\ 
&f_7(d;u) = (d-4)^2\, M_7(d;u)\,, \nonumber\\ 
&f_8(d;u) = (d-4)^3\, (d-3)\,u\, M_8(d;u)\,.  \labbel{eq:basis1}
\end{align}
The system of differential equations for the first five masters integrals 
can then be written as
\begin{align}
\frac{d}{du} f_i(d;u) = (d-4)\, \sum_{j=1}^5\, A_{ij}(u)\, f_j(d;u)\,, 
                            \quad \forall i=1,...,5 \labbel{eq:deqca}
\end{align}
where the matrix $A(u)$ reads
\begin{align}
A(u) = \frac{1}{u}\; \left( \begin{array}{ccccc} 
0 & 0 & 0 & 0 & 0 \\
0 & -1/2 & 0 & 0 & 0 \\
0 & 0 & -1/2 & 0 & 0 \\
0 & 0 & 3/2 & 0 & 0 \\
0 & 0 & 0 & 0 & -1
 \end{array}   \right)\, 
+ \frac{1}{u-1}\; \left( \begin{array}{ccccc} 
0 & 0 & 0 & 0 & 0 \\
1/8 & 1 & 0 & 0 & 0 \\
0 & 0 & 0 & -1 & 0 \\
0 & 0 & 0 & 2 & 0 \\
0 & 1 & 0 & 0 & 2 
\end{array}   \right)\,. \labbel{eq:matrix}
\end{align}

The differential equations for the last three integrals cannot be put in 
a similarly simple form
and we will write them explicitly later on, once we come to study them.

\section{The simple kite master integrals} 
\labbel{sec:simple} \setcounter{equation}{0} 
\numberwithin{equation}{section}
Let us focus on the first five integrals. If we start from the differential 
equations~\eqref{eq:deqca}, carrying out the integration
in terms of harmonic polylogarithms is straightforward. As it is well known, harmonic polylogarithms are
a special case of multiple polylogarithms and 
for convenience of the reader, we recall here their iterative definition.
We start at weight one defining
\begin{align}
G(0,x) = \ln{(x)}\,, \qquad G(a,x) = \int_0^x \frac{dt}{t-a} = \ln{\left( 1 - \frac{x}{a}\right)}\,.
\end{align}
The multiple polylogarithms  are then iteratively defined at weight $n$ as follows
\begin{align}
G(\underbrace{0,...,0}_{n};x) = \frac{1}{n!} \ln^n{x}\,, \qquad 
G(a_1,a_2,...,a_n; x) = \int_0^x \frac{dt}{t-a_1} G(a_2,...,a_n; x)\,.
 \labbel{eq:defMPLs}
\end{align}

Note that all integrals have a cut at $s=m^2$, $u=1$, {\it i.e.} they are real for $u<1$ and
develop an imaginary part for $u>1$ whose sign is fixed by Feynman's prescription $u \to u + i\, 0^+$.
We present here the solution valid for for $0<u<1$. The analytic continuation
to the physical region can be then easily obtained by continuing to $u>1$ with $u \to u + i\, 0^+$.
Thanks to the choice of a canonical basis the solution takes a particularly compact form and all
sub-topologies, up to weight 4, fit in one single page.

\begin{align}
f_1(d;u) &= 1\,,  \label{f1du} \\  \nonumber \\
f_2(d;u) &= \frac{(d-4)}{8}\, G(1,u) + \frac{(d-4)^2}{8} \left[ G(1,1,u) - \frac{1}{2} G(0,1,u) \right] \nonumber \\
&+\frac{(d-4)^3}{8} \left[ G(1,1,1,u) - \frac{1}{2}G(1,0,1,u) - \frac{1}{2} G(0,1,1,u) + \frac{1}{4}G(0,0,1,u) \right]
\nonumber \\
&+ \frac{(d-4)^4}{8} \left[ G(1,1,1,1,u) - \frac{1}{2}G(1,1,0,1,u) - \frac{1}{2}G(1,0,1,1,u) 
+ \frac{1}{4}G(1,0,0,1,u)
 \right. \nonumber \\
&\qquad \qquad \; \left. -\frac{1}{2}G(0,1,1,1,u) + \frac{1}{4}G(0,1,0,1,u) + \frac{1}{4} G(0,0,1,1,u) - \frac{1}{8}G(0,0,0,1,u)
\right]\,,\\  \nonumber \\
f_3(d;u) &= - \frac{(d-4)}{8}\, G(1,u) + \frac{(d-4)^2}{4} \left[ - G(1,1,u) + \frac{1}{4} G(0,1,u) \right] \nonumber \\
&+\frac{(d-4)^3}{2} \left[ - G(1,1,1,u) + \frac{3}{8}G(1,0,1,u) - \frac{\pi^2}{48} G(1,u)
+ \frac{1}{4} G(0,1,1,u) - \frac{1}{16}G(0,0,1,u) \right] 
\nonumber \\
&+ (d-4)^4 \left[ -G(1,1,1,1,u) + \frac{3}{8}G(1,1,0,1,u) - \frac{\pi^2}{48} G(1,1,u) + \frac{3}{8}G(1,0,1,1,u) 
 \right. \nonumber \\
&\qquad \qquad \; - \frac{3}{32}G(1,0,0,1,u)
 -\frac{\zeta_3}{32} G(1,u) + \frac{1}{4}G(0,1,1,1,u) - \frac{3}{32}G(0,1,0,1,u)  \nonumber \\
&\qquad \qquad \; \left.
+ \frac{\pi^2}{192} G(0,1,u)
- \frac{1}{16} G(0,0,1,1,u) + \frac{1}{64}G(0,0,0,1,u)
\right]\,, \\  \nonumber \\
f_4(d;u) &= \frac{1}{8} + \frac{(d-4)}{4}\, G(1,u) + \frac{(d-4)^2}{2} \left[ G(1,1,u) - \frac{3}{8} G(0,1,u) + \frac{\pi^2}{48} 
\right] \nonumber \\
&+(d-4)^3 \left[ G(1,1,1,u) - \frac{3}{8}G(1,0,1,u) +\frac{\pi^2}{48} G(1,u)
- \frac{3}{8} G(0,1,1,u) + \frac{3}{32}G(0,0,1,u) + \frac{\zeta_3}{32} \right]
\nonumber \\
&+ (d-4)^4 \left[ 2\,G(1,1,1,1,u) - \frac{3}{4}G(1,1,0,1,u) + \frac{\pi^2}{24} G(1,1,u) - \frac{3}{4}G(1,0,1,1,u) 
 \right. \nonumber \\
&\qquad \qquad \; + \frac{3}{16}G(1,0,0,1,u)
 + \frac{\zeta_3}{16} G(1,u) - \frac{3}{4}G(0,1,1,1,u) + \frac{9}{32}G(0,1,0,1,u)  \nonumber \\
&\qquad \qquad \; \left.
- \frac{\pi^2}{64} G(0,1,u)
+ \frac{3}{16} G(0,0,1,1,u) - \frac{3}{64}G(0,0,0,1,u) + \frac{\pi^4}{1280}
\right]\,,\\  \nonumber \\
f_5(d;u) &= \frac{(d-4)^2}{8} G(1,1,u) 
+ \frac{(d-4)^3}{8} \left[ 3\,G(1,1,1,u) - \frac{1}{2}G(1,0,1,u) + G(0,1,1,u) \right] \nonumber \\
&+ \frac{(d-4)^4}{8} \left[ 7\, G(1,1,1,1,u) - \frac{3}{2} G(1,1,0,1,u) - \frac{5}{2} G(1,0,1,1,u) + \frac{1}{4} G(1,0,0,1,u) \right.
\nonumber \\
&\qquad \qquad \; \left. -3\,G(0,1,1,1,u) + \frac{1}{2} G(0,1,0,1,u) + G(0,0,1,1,u)
 \right]\,. \labbel{eq:subeasy}
\end{align}

\section{The choice of the basis for the sunrise amplitudes.} 
\labbel{sec:basis} \setcounter{equation}{0} 
\numberwithin{equation}{section}
We move now to consider the last three integrals. First of all we need to 
focus on the two master
integrals of the two-loop sunrise graph, {\it i.e.} $f_6(d;u)$ and $f_7(d;u)$. 
They satisfy a system of two coupled differential equations
\begin{align}
 u\,\frac{d}{du} f_6(d;u) & = -f_6(d;u) + 3f_7(d;u) 
                                 + (d-2)f_6(d;u)\,,\nonumber \\&\nonumber \\
 u(u-1)(u-9)\,\frac{d}{du} f_7(d;u) & = (u-3)f_6(d;u) 
                                 - (u^2-9)f_7(d;u) \nonumber \\
                          & + (d-2)\left[ -\frac{5}{2}(u-3)f_6(d;u) 
                   +\frac{u^2+10 u -27}{2} f_7(d;u)\right]\nonumber \\
                          & + (d-2)^2 \frac{3\,(u-3)}{2}f_6(d;u)  
                   -   \,\frac{u}{2}\,f_1(d;u)\,. \labbel{eq:deq1}
\end{align} 
Let us recall here that amplitude $ f_1(d;u) $ appearing within the 
inhomogeneous term corresponds to the product of two tadpoles and is 
in fact constant, according to Eq.(\ref{f1du}). \\ 
Using the methods described in~\cite{Tancredi:2015pta} one can show
that it is not possible to decouple the system,
in any even number of dimensions $d=2\,n$, $n \in \mathbb{N}$, by taking 
simple linear combinations of the masters integrals with rational 
coefficients. Indeed, it is very well known that the solution cannot be 
expressed in terms of MPLs only and elliptic generalizations of the
latter must be introduced~\cite{Laporta:2004rb,Bloch:2013tra,Remiddi:2013joa,
Adams:2013nia,Adams:2014vja,Adams:2015gva}. 

In order to simplify the integration of these two integrals
we will proceed as follows. We will start considering the integrals for 
$d \approx 2$. The reason for this is two-fold. First, when $d=2$ the two 
master integrals $f_6(2;u)$ and $f_7(2;u)$ are finite. 
Second, as we will see explicitly, their imaginary parts when $d=2$ are 
particularly simple. That is important because the imaginary parts are 
related to the solutions of the corresponding homogeneous system, which 
in turn are the building blocks for the iterative solution of the 
$2 \times 2$ differential system~\eqref{eq:deq1} through Euler's 
method. Those considerations will allow us 
to determine a basis of master integrals for which we can easily solve 
the differential equations as a Laurent series in $(d-2)$. 
At this point, we could solve the system as Laurent series in $(d-2)$ and then 
use Tarasov shifting identities~\cite{Tarasov:1996br}
in order to obtain the corresponding coefficients of their Laurent series 
in $(d-4)$, which are the physically relevant results.
Instead of proceeding in this way, though, we will use the technique
described in~\cite{Tancredi:2015pta} (see appendix B therein) in order to 
build up a \textsl{new basis} of master integrals which fulfills the very 
same differential equations, but this time with $d \to d-2$. 
This implies that the series expansion in $(d-4)$ of the new basis will be 
\textsl{formally identical} to that of the former basis in $(d-2)$.
This will allow us to treat more consistently everything in $d\approx 4$ 
from the very beginning.

\subsection{Simplifying the differential equations in $d=2$}
The system of differential equations~\eqref{eq:deq1} has four regular 
singular points, {\it i.e.} $u=0$, $u=1$, $u=9$ and $u = \pm \infty$, we will 
therefore need to consider the solution in the four different regions 
$$-\infty<u<0\,,\quad 0<u<1\,,\quad 1<u<9\,,\quad 9<u<\infty\,.$$ 
Physically, the point $u=9$, $s = 9\,m^2$, corresponds to the three massive 
particle cut and we expect the master integrals to develop an imaginary part 
as $u>9$.
Now, since the tadpole does not have any cut in $u$, the imaginary parts 
of the master integrals $f_6(d;u)$ and $f_7(d;u)$
must satisfy the associated homogeneous system. 
We have already computed the imaginary part of the first master integral in 
section~\ref{sec:iter} for generic $d$. 
For $d=2$, a straightforward application of Cutkosky-Veltman's rule to 
$ f_7(2;u) $ as well gives 
\begin{align} 
 & \frac{1}{\pi} \Ima f_6(2;u) = I(0,u) \nonumber \\
 & \frac{1}{\pi} \Ima f_7(2;u) = \frac{1}{(u-1)(u-9)} 
                                \left[ \frac{u^2-6\,u + 21}{6}\,I(0,u) 
 - \frac{1}{2} I(2,u) \right]\,,\labbel{eq:impartseq} \\&\nonumber
\end{align}
where the functions $I(n,u)$ are defined as 
\begin{align}
&I(n,u) = \int_4^{(\sqrt{u}-1)^2} db\, \frac{b^n}{\sqrt{R_4(b,u)}}\,,
\labbel{defInu} \end{align}
and $R_4(d,u)$ is the fourth-order polynomial
\begin{align}
R_4(b,u) = b(b-4)((\sqrt{u}-1)^2-b)((\sqrt{u}+1)^2-b)\,. \labbel{defR4} 
\end{align}
Some of the properties of these functions are discussed
in appendices~\ref{App:Ell} and~\ref{App:CEllInt}.
In particular, there it is shown that all functions $I(n,u)$ can be 
expressed in terms of two independent functions only, say $I(0,u)$ 
and $I(2,u)$, which can be in turn expressed in terms of the complete 
elliptic integrals of first and second kind, and can be therefore considered 
as known analytically. Eqs.~\eqref{eq:impartseq} suggest to perform 
the following change of basis
\begin{align}
 g_6(d;u) &= f_6(d;u) \nonumber \\
 g_7(d;u) &= - 2 (u-1) (u-9) f_7(d;u) + \frac{1}{3}( u^2 - 6 \,u + 21 ) 
                f_6(d;u)\,,\labbel{eq:transf1}
\end{align}
so that obviously the imaginary parts of the functions 
$g_6(d;u)$ and $g_7(d;u)$  in $d=2$ become
\begin{align} 
 \frac{1}{\pi} \Ima \,g_6(2;u) &= I(0,u)\,,\nonumber\\ 
 \frac{1}{\pi} \Ima \,g_7(2;u) &= I(2,u)\,. \labbel{Img67} 
\end{align} 
Note that these relations are true in $d=2$ and do not change if we 
modify~\eqref{eq:transf1} by a term proportional to $(d-2)$. 
This freedom can be used to get rid of the term
proportional to $(d-2)^2$ in the second of Eqs.~\eqref{eq:deq1}.
While this is not strictly required, it indeed helps in simplifying
the structure of the solution.
We modify therefore Eq.\eqref{eq:transf1} as follows
\begin{align}
 g_6(d;u) &= f_6(d;u) \nonumber \\
 g_7(d;u) &= - 2 (u-1) (u-9) f_7(d;u) + \frac{1}{3}( u^2 - 6 \,u + 21 ) 
              f_6(d;u) + (d-2) C(u) f_6(d;u)\,,\labbel{eq:transf2}
\end{align}
where $C(u)$ is a function of $u$ only, to be determined by imposing that 
the term proportional to $(d-2)^2$ in Eqs~\eqref{eq:deq1} is zero. 
By writing down explicitly the differential equations for $g_6(d;u)$ and 
$g_7(d;u)$ one easily finds that there are two values of $C(u)$ which would 
eliminate the unwanted term, namely
$$C(u) = 6(u-1)\,,\qquad C(u) = -\frac{(u-3)(u-9)}{3}\,.$$
At this level, there is no reason to prefer one choice over the other, 
we choose therefore $C(u) = 6(u-1)$, since this produces the most compact 
results. With this choice the differential equations become
\begin{align}
 \frac{d}{du} g_6(d;u) & = \frac{1}{2 u (u-1)(u-9)} \left[ 
               (3+14 u - u^2)g_6(d;u) - 3\,g_7(d;u) \right]
  + (d-2)\,\frac{1}{u-9}\,g_6(d;u)\,,\nonumber \\&\nonumber \\
 \frac{d}{du} g_7(d;u) & = \frac{1}{6\, u (u-1)(u-9)} 
                       \left[ (u+3)(3+75u-15u^2+u^3)g_6(d;u) 
                             - 3(3+14u-u^2) g_7(d;u)\right]\nonumber \\
                       & + \frac{(d-2)}{6\, u(u-1)(u-9)} 
                       \left[ (u+3)(9+63u-9u^2+u^3)g_6(d;u) 
                              + 3(u+1)(u-9)g_7(d;u) \right]
                       + 1\,, \labbel{eq:deq2}
\end{align} 
where we used $f_1(d;u) = 1$, Eq.(\ref{f1du}). 
The system can be written in matrix form as follows

\begin{align} 
 \frac{d}{du} \left( \begin{array}{c} g_6\\g_7 \end{array} \right) 
 = B(u) \left( \begin{array}{c} g_6\\g_7 \end{array} \right) 
 + (d-2) \, D(u)\, \left( \begin{array}{c} g_6\\g_7 \end{array} \right) 
 + \left( \begin{array}{c} 0\\1 \end{array} \right)\,, \labbel{eq:sysd2}
\end{align}
where the two matrices $ B(u), D(u) $ are defined as
\begin{align}
 B(u) = \frac{1}{6\, u (u-1)(u-9)} 
 \left( \begin{array}{cc} 3(3+14u-u^2) & - 9  \\ (u+3)(3+75u-15u^2+u^3)  
             & - 3(3+14u-u^2) \end{array} \right)\,,
\end{align}
\begin{align}
 D(u) = \frac{1}{6\,u(u-9)(u-1)} 
 \left( \begin{array}{cc} 6\,u(u-1) & 0  \\ 
 (u+3)(9+63u-9u^2+u^3)  & 3(u+1)(u-9) \end{array} \right)\,.
\end{align}

In order to be able to solve~\eqref{eq:sysd2} 
as a Laurent series in $(d-2)$, as a first step we need to solve the 
homogeneous system for $d=2$, {\it i.e.} we need to find a pair of two solutions,
say $(I_1(u),I_2(u))$ and $(J_1(u), J_2(u))$, such that the matrix of the 
solutions
\begin{align}
 G(u) = \left( \begin{array}{cc} I_1(u) & J_1(u) \\ I_2(u) & J_2(u) 
               \end{array} \right) \labbel{defG} 
 \end{align} 
fulfills 
 \begin{align}
 \frac{d}{du} G(u) = B(u)\, G(u)\,.  \labbel{eqG} 
\end{align}
Note in particular that since 
 \begin{equation} 
      {\rm Tr}{(B(u))} = 0\,\,, 
 \end{equation} 
the Wronskian of the four solutions, 
$W(u) = I_1(u)\,J_2(u) - I_2(u)\,J_1(u)$, must be independent of $u$.
From its very definition, 
\begin{align}
 W(u) = \det{(G(u))} = I_1(u)\,J_2(u) - I_2(u)\,J_1(u)\,, \labbel{eq:Wronsk}
\end{align}
we find
\begin{align}
 \frac{d}{du} W(u) = \frac{d}{du} \det{(G(u))} 
                   = {\rm Tr}(G^{-1}(u)\,B(u)\,G(u))\, \det{(G(u))}
                    = {\rm Tr}(B(u))\, \det{(G(u))} = 0\,,
\end{align}
and $W(u)$ must be a constant. This property is of fundamental importance 
to simplify the iterative solution of the differential equations, as we 
will see later on.

\subsection{The choice of the basis for the expansion in $(d-4)$. }
In the previous section we showed how to choose a basis of master integrals 
for the sunrise graph, whose differential equations take a particularly 
convenient form as far as their Laurent series in $(d-2)$ are considered. 
Since, as it is well known, any Feynman integral in $d-2$ dimensions can 
be expressed as a linear combination of Feynman integrals in $d$ dimensions, 
by following the method described in appendix B of~\cite{Tancredi:2015pta} 
we define a new basis of master integrals by shifting~\eqref{eq:transf2} 
from $d \to d-2$ 
\begin{align}
h_6(d;u) = g_6(d-2,u)\,, \qquad h_7(d,u) = g_7(d-2,u)\,.
\end{align}
Using Tarasov's relations we find that the new basis $h_6(d;u),\, h_7(d;u)$ 
can be written in terms of the original master integrals $f_6(d;u)$ and 
$f_7(d;u)$ (and their sub-topology $f_1(d,u)$) as follows

\begin{align}
h_6(d;u) &=  \frac{12\,(d-3)\,(3\,d-8)}{(u-1)(u-9)}\,f_6(d;u) 
               + \frac{24\,(d-3)(u+3)}{(u-1)(u-9)} \,f_7(d;u)
- \frac{3 \,(u-3) }{(u-1)(u-9)}\,f_1(d;u)\,,
\nonumber \\ & \nonumber \\
h_7(d;u) &= \frac{4(d-3)(3\,d-8)(3 - (58-18 \, d)\,u 
       + (7-2\,d)\,u^2)}{(u-1)(u-9)} \, f_6(d;u) \nonumber \\
&+ \frac{8\,(d-3) ( 9 + 9 (9 \, d - 29)\, u - 9 (2\,d-7)\, u^2 
                     + (d-3)\, u^3)}{(u-1)(u-9)}\, f_7(d;u) \nonumber \\
&+ \frac{ (9 - (51 - 18\, d) \, u - (61 - 16\, d) \,u^2 
      + (7 - 2\, d) \,u^3)}{(u-1)(u-9)}\, f_1(d;u)\,. \labbel{eq:transf3}
\end{align}
It is straightforward by direct calculation, and using~\eqref{eq:sysd2}, 
to prove that the new basis~\eqref{eq:transf3} satisfies
the new system of differential equations
\begin{align} 
 \frac{d}{du} \left( \begin{array}{c} h_6\\h_7 \end{array} \right) 
 = B(u) \left( \begin{array}{c} h_6\\h_7 \end{array} \right) 
 + (d-4) \, D(u)\, \left( \begin{array}{c} h_6\\h_7 \end{array} \right) 
 + \left( \begin{array}{c} 0\\1 \end{array} \right)\,. \labbel{eq:sysd4}
\end{align}
As expected the system~\eqref{eq:sysd4} is identical to~\eqref{eq:sysd2}, 
upon the formal substitution $d \to d-2$. This also implies that all the
properties fulfilled by $g_6(d;u)$ and $g_7(d,u)$ in the limit $d \to 2$, 
are also fulfilled by $h_6(d;u)$ and $h_7(d;u)$ in the limit 
$d \to 4$. In particular the new master integrals are \textsl{finite} 
in $d=4$ and their imaginary parts read
\begin{align} 
 \frac{1}{\pi} \Ima \,h_6(4;u) &= I(0,u)\,, \nonumber\\ 
 \frac{1}{\pi} \Ima \,h_7(4;u) &= I(2,u)\,, \labbel{Imh67} 
\end{align} 
as Eqs.(\ref{Img67}).
\section{The solution of the differential equations} 
\labbel{sec:sun} \setcounter{equation}{0} 
\numberwithin{equation}{section}
In this section we will show how to build the complete solution for 
the sunrise master integrals up to any order in $(d-4)$. We will solve 
the system as Laurent series in $(d-4)$.
In order to do this, we first need to find the homogeneous solution in the 
limit $d \to 4$, such that we can then use Euler's method of variation of 
constants in order to build up the complete non-homogeneous solution.

\subsection{The homogeneous solution}

As explained above, as a first step we need now to find two independent 
pairs of solutions for the homogeneous system associated 
to~\eqref{eq:sysd4} 
\begin{equation}
 \frac{d}{du} \left( \begin{array}{c} I_1 \\ I_2 \end{array} \right) 
       = B(u) \left( \begin{array}{c} I_1 \\ I_2 \end{array} \right) \,. 
 \labbel{eq:homd4} 
\end{equation} 
The discussion in previous section already suggests how to find the first 
of the two pairs. Taking the imaginary part of~\eqref{eq:sysd4} at 
$ d=4 $ gives at once 
\begin{equation}
 \frac{d}{du} \left( \begin{array}{c} \Ima\,h_6\\ \Ima\, h_7 
                                                  \end{array} \right) 
 = B(u) \left( \begin{array}{c} \Ima\, h_6\\ \Ima\, h_7 \end{array} 
                                       \right) \,, \labbel{eq:imhomd4}
\end{equation}
so that Eqs.(\ref{Imh67}) provide obviously with a first pair of solution, 
valid for $9<u<\infty$, 
\begin{align}
I_1^\noo(u) &= I(0,u)\,, \nonumber\\ 
I_2^\noo(u) &= I(2,u)\,. \labbel{ImI12} 
\end{align}
It is straightforward, using the results of appendix~\ref{App:Ell}, to 
compute the derivatives of these functions, obtaining 
\begin{align}
 \frac{d}{du} I(0,u) & = \frac{1}{2 u (u-1)(u-9)} 
 \left[ (3+14 u - u^2)I(0,u) - 3\,I(2,u) \right]\,,
 \nonumber \\
 \frac{d}{du} I(2,u) & = \frac{1}{6\, u (u-1)(u-9)} 
                       \left[ (u+3)(3+75u-15u^2+u^3)I(0,u) 
                             - 3(3+14u-u^2)I(2,u)\right]\,,\labbel{derIm} 
\end{align} 
which can be also written as 
\begin{align} 
\frac{d}{du} \left( \begin{array}{c}  I_1^\noo(u) \\ I_2^\noo(u) 
                                            \end{array}\right) = 
B(u)\, \left( \begin{array}{c}  I_1^\noo(u) \\ I_2^\noo(u) 
                   \end{array}\right)\,, 
\end{align} 
as expected. 

In order to find a second pair of solutions, we go back to the definition 
of the functions $I(n,u)$, introduced in Eq.(\ref{defInu}) as the definite 
integral in $b$ of the square root of the fourth-order polynomial 
$R_4(b,u)$, Eq.(\ref{defR4}), between two adjacent roots. 
Since $R_4(b,u)$ has 4 roots, we are naturally brought to consider two 
similar sets of functions, defined by integrating between the other 
pairs of adjacent roots, say
\begin{align}
&J(n,u) = \int_0^4 db\, \frac{b^n}{\sqrt{-R_4(b,u)}}\,, 
          \hspace{1cm} {\rm or} 
&K(n,u) = \int_{(\sqrt{u}-1)^2}^{(\sqrt{u}+1)^2} db\, 
\frac{b^n}{\sqrt{-R_4(b,u)}}\,. \labbel{defJK} 
\end{align}
More details on these functions are provided in appendix~\ref{App:Ell}. 
In particular, one can show that, also in this case, there are two 
``master integrals" for each set of functions, say $J(0,u)$, $J(2,u)$ 
and $K(0,u)$, $K(2,u)$. Moreover, one can show that the functions $K(n,u)$ 
are not independent from the functions $J(n,u)$ and we can therefore 
neglect them. We pick for definiteness $J(0,u)$ and $J(2,u)$ and we 
compute their derivatives finding
\begin{align}
 \frac{d}{du} J(0,u) = \frac{1}{2 u (u-1)(u-9)} 
 &\left[ (3+14 u - u^2)J(0,u) - 3\,J(2,u) - \pi (u+3) \right]\,,
 \nonumber \\
 \frac{d}{du} J(2,u) = \frac{1}{6\, u (u-1)(u-9)} 
              &\left[ (u+3)(3+75u-15u^2+u^3)J(0,u) 
                - 3(3+14u-u^2)J(2,u)\right. \nonumber \\
              &  \left. - \pi (9 + 63u -9u^2+u^3) \right]\,.\labbel{derJ}
\end{align} 
$J(0,u)$ and $J(2,u)$, as they stand, are not solutions of the 
homogeneous system; it is nevertheless very easy to use them in order to 
build a proper solution for the system. Consider the new functions 
defined as
\begin{align*}
 \bar{J}(0,u) = J(0,u)\,,\qquad \bar{J}(2,u) = J(2,u) + \frac{\pi}{3}(u+3).
\end{align*}
By using~\eqref{derJ} it is trivial to verify that their derivatives read
\begin{align}
 \frac{d}{du} \bar{J}(0,u) = \frac{1}{2 u (u-1)(u-9)} 
 &\left[ (3+14 u - u^2)\bar{J}(0,u) - 3\,\bar{J}(2,u) \right]\,,
 \nonumber \\
 \frac{d}{du} \bar{J}(2,u) = \frac{1}{6\, u (u-1)(u-9)} 
                       &\left[ (u+3)(3+75u-15u^2+u^3)\bar{J}(0,u) 
                     - 3(3+14u-u^2)\bar{J}(2,u) \right]\,,\labbel{derJb}
\end{align}
so that we can define our second set of solutions, again for $9<u<\infty$, as
\begin{align}
J_1^\noo(u) &= \int_0^4 \frac{db}{\sqrt{-R_4(b,u)}}\,, \nonumber\\ 
J_2^\noo(u) &= \int_0^4 \frac{db\ b^2}{\sqrt{-R_4(b,u)}}\, + \frac{\pi}{3}(u+3) \,. 
\labbel{defJ12} 
\end{align}

Summarising we have found two pairs of independent 
real valued solutions, valid in the range $9<u<\infty$, such that their matrix 
\be G^\noo(u)= 
\left( \begin{array}{cc} I_1^\noo(u) & J_1^\noo(u) \\ 
                         I_2^\noo(u) & J_2^\noo(u) \end{array}\right) 
\labbel{defG9oo} \ee 
fulfils 
\be \frac{d}{du} G^\noo(u) = B(u)\, G^\noo(u)\,. \labbel{eqG9oo} \ee 

We can now proceed and study their limiting behaviour on the two boundaries,
{\it i.e.} $u \to 9^+$ and $u \to +\infty$.
For $u \to 9^+$ we find (keeping only the leading logarithmic behaviour)
\begin{align}
 &I_1^{\noo}(u\to9^+) = \frac{\sqrt{3}}{12} \pi\,, \nonumber\\ 
 &I_2^{\noo}(u\to9^+) = \frac{4\sqrt{3}}{3} \pi\,, \nonumber\\ 
 &J_1^{\noo}(u\to9^+) =   
 \frac{\sqrt{3}}{2 } \left( \frac{\ln{3}}{3} + \frac{\ln{2}}{2} 
                              - \frac{\ln{(u-9)}}{6}\right)\,, \nonumber\\ 
 &J_2^{\noo}(u\to9^+) = 
  4 \sqrt{3} \left( 1 + \frac{2\,\ln{3}}{3} + \ln{2} 
                                     - \frac{\ln{(u-9)}}{3}\right) \,.
\end{align} 
On the other hand for $u \to + \infty$ we find
\begin{align}
 &I_1^{\noo}(u\to+\infty) = \frac{3}{2} \frac{\ln{(u)}}{u}\,,\qquad 
 I_2^{\noo}(u\to+\infty)  = \frac{1}{2}\,u\,\ln{(u)} -u\,, \nonumber\\
 &J_1^{\noo}(u\to+\infty) =   \frac{\pi}{u}\,,\qquad \qquad
 J_2^{\noo}(u\to+\infty) = \frac{\pi}{3} u\,.
 \end{align}
As stated previously, the Wronskian~\eqref{eq:Wronsk} must be independent 
of $u$; when can computing it using any of the limits above, we find 
\begin{equation} 
\lim_{u \to 9^+}W(u) 
= \lim_{u \to 9^+} \left( I_1^{\noo}(u )J_2^{\noo}(u ) 
                               - I_2^{\noo}(u )  J_1^{\noo}(u ) \right)  
= \pi, \labbel{eq:valWronsk1}
\end{equation}
and 
\begin{equation} \lim_{u \to + \infty }W(u) 
= \lim_{u \to +\infty} \left( I_1^{\noo}(u )J_2^{\noo}(u) 
           - I_2^{\noo}(u)  J_1^{\noo}(u) \right)  
= \pi, \labbel{eq:valWronsk2} 
\end{equation} 
as expected. 
\newline 

The solution described here is valid above threshold, {\it i.e.} for 
$ u > 9 $, but it is straightforward to extend those results and build up 
a complete set of solutions valid in the remaining regions, i.e. 
$-\infty<u<0$, $0<u<1$ and $1<u<9$. The details are worked out explicitly in 
appendix~\ref{App:AnCont}. We end up in this way with 4 different matrices 
of \textsl{real} solutions $G^{(a,b)}(u)$, each valid in the interval $a<u<b$,
and which can be continued from one region to the other using the matching 
matrices given in the same appendix, see in particular 
Eqs.~\eqref{eq:matching1} and~\eqref{eq:matching2}.
Note that in~\eqref{eq:valWronsk1} and~\eqref{eq:valWronsk2} we computed 
the value of the Wronskian in the region $9<u<\infty$, but we can normalize the solutions
in the remaining three regions such that the same remains true in every interval $(a,b)$, see
appendix~\ref{App:AnCont},
\begin{equation}
I^{(a,b)}_1(u)J^{(a,b)}_2(u) - I^{(a,b)}_2(u) J^{(a,b)}_1(u) = \pi\,. 
\labbel{eq:Wronsknum}
\end{equation}

\subsection{The non-homogeneous solution}
Once we have the homogeneous solution of the system for $d=4$ we can use 
Euler's method of the variation of constants in order to write the complete 
solution of the system Eq.(\ref{eq:sysd4}). The manipulations performed 
here are the same for all the regions $a<u<b$, 
we will therefore drop the superscripts $(a,b)$ from all formulas 
for simplicity, writing for instance $ G(u) $ instead of $ G^\noo(u) $ 
etc. It will be then simple to specialize the results to the 
region of interest by picking the suitable set of solutions $G^{(a,b)}(u)$, 
see for instance Eqs.(\ref{defG9oo},\ref{eqG9oo}) for the notation. 
We perform the rotation
\begin{align}
 \left( \begin{array}{c} h_6(d;u)\\h_7(d;u) \end{array} \right) = G(u)\, 
 \left( \begin{array}{c} m_6(d;u)\\m_7(d;u) \end{array} \right)\,, 
 \labbel{eq:rotation}
\end{align}
such that the new functions $m_6(d;u)$ and $m_7(d;u)$ fulfil the equations
\begin{align*}
 \frac{d}{du} \left( \begin{array}{c} m_6(d;u)\\m_7(d;u) \end{array} \right)
 = (d-4)\, G^{-1}(u)\, D(u)\, G(u) \, 
   \left( \begin{array}{c} m_6(d;u)\\m_7(d;u) \end{array} \right)
   +  G^{-1}(u)\left( \begin{array}{c} 0 \\ 1 \end{array} \right)\,.
\end{align*}
Thanks to the condition on the Wronskian~\eqref{eq:Wronsknum},
inverting the matrix $G(u)$ is straightforward and we get
\begin{align}
 G^{-1}(u) = \frac{1}{\pi}\,\left( \begin{array}{cc} J_2(u) & -J_1(u) \\ 
                -I_2(u) & I_1(u) \end{array} \right)\,.
\end{align}
We write therefore the system as
\begin{align}
 \frac{d}{du} \left( \begin{array}{c} m_6(d;u)\\m_7(d;u) \end{array} \right)
 = (d-4)\, \frac{1}{\pi} M(u) \, \left( \begin{array}{c} m_6(d;u)\\ 
                           m_7(d;u) \end{array} \right)
 +  \frac{1}{\pi} \left( \begin{array}{c} -J_1(u) \\ 
                  I_1(u) \end{array} \right)\,, \labbel{eq:systosolve}
\end{align}
where we introduced the matrix
\begin{align}
 M(u) = \pi\, G^{-1}(u)\, D(u)\, G(u)\,. 
\end{align} 
Written in this form, the iterative structure of the solution in powers 
of $(d-4)$ becomes manifest. \par 
The entries of the matrix $M(u)$ read
\begin{align}
 M_{11}(u) &= + \frac{I_1(u)J_2(u)}{u-9} 
              - \frac{(u+1)I_2(u)J_1(u)}{2\,u\,(u-1)}
            - \frac{(u+3)\left[ \,9+\,u\,(63+(u-9)u)\, \right] 
                           I_1(u)J_1(u)}{6\,u\,(u-1)\,(u-9)}\,, 
           \nonumber&\\
 M_{12}(u) &= - \frac{(u+3)}{6\,u\,(u-1)\,(u-9)} 
              \left\{ \left[ \,9+\,u\,(63+(u-9)u)\, \right]J_1^2(u) 
                      -  3(u+3) J_1(u)\,J_2(u) \right\} \,, \nonumber&\\
 M_{21}(u) &= + \frac{(u+3)}{6\,u\,(u-1)\,(u-9)} 
              \left\{ \left[ \,9+\,u\,(63+(u-9)u)\, \right]I_1^2(u) 
                      -  3(u+3) I_1(u)\,I_2(u) \right\} \,,\nonumber &\\   
 M_{22}(u) &= -\frac{I_2(u)J_1(u)}{u-9} 
              + \frac{(u+1)I_1(u)J_2(u)}{2\,u\,(u-1)}
            + \frac{(u+3)\left[ \,9+\,u\,(63+(u-9)u)\, \right] 
                           I_1(u)J_1(u)}{6\,u\,(u-1)\,(u-9)}\,; 
\labbel{Matrix} 
\end{align} 
they contains rational functions and products of pairs of homogeneous 
solutions, i.e. products of \textsl{complete elliptic integrals}. 
\par 
It should be recalled at this point that not all products are actually 
linearly independent; because of the condition on the Wronskian, in fact, 
only one of the two combinations $I_1(u)J_2(u)$ or $I_2(u) J_1(u)$ is 
really independent, while the other can be removed using~\eqref{eq:Wronsknum}.
Moreover, notice that the functions $I_k(u)$ and $J_k(u)$ fulfil the same 
differential equations, i.e.~\eqref{derIm} or~\eqref{derJb}. 
By inverting them one can, for example, get rid of $I_2(u)$
and $J_2(u)$ in favour of $I_1(u)$ and $J_1(u)$ and their derivatives
\begin{align}
&I_2(u) = -\frac{(u^2 - 14\, u - 3)}{3}\, I_1(u) 
          - \frac{2}{3}\,u\,(u-1)\,(u-9)\, \frac{d\, I_1}{d\,u}\,,\\
&J_2(u) = -\frac{(u^2 - 14\, u - 3)}{3}\, J_1(u) 
          - \frac{2}{3}\,u\,(u-1)\,(u-9)\, \frac{d\, J_1}{d\,u}\,.
\end{align}
Substituting these relations into~\eqref{Matrix} and rearranging the terms, 
the matrix can be written in a much more compact form as 
\begin{align} 
 M_{11}(u) &= - \frac{d}{d\,u}\left( \frac{(u+3)^2}{6} I_1(u)\, J_1(u)  
                              \right) 
              + \frac{\pi}{4} \left( \frac{2}{u-9}  + \frac{2}{u-1} 
              - \frac{1}{u} \right)\,, \nonumber&\\
 M_{12}(u) &= - \frac{d}{d\,u}\left( \frac{(u+3)^2}{6} I_1(u)\, I_1(u) 
                              \right) \,, \nonumber&\\
 M_{21}(u) &= + \frac{d}{d\,u}\left( \frac{(u+3)^2}{6} J_1(u)\, J_1(u) 
                              \right)\,, \nonumber &\\   
 M_{22}(u) &= + \frac{d}{d\,u}\left( \frac{(u+3)^2}{6} I_1(u)\, J_1(u) 
                              \right) 
              +  \frac{\pi}{4} \left( \frac{2}{u-9}  + \frac{2}{u-1} 
              - \frac{1}{u} \right)\,, \labbel{Matrix3} 
\end{align} 
where we used 
\begin{equation} 
 I_1(u)\, \frac{d\,J_1(u)}{d u} - J_1(u)\, \frac{d\,I_1(u)}{d u} 
                           = - \frac{3\,\pi}{2} \frac{1}{u(u-1)(u-9)}\,,
\end{equation}
which can be easily proved starting from the condition on the 
Wronskian~\eqref{eq:Wronsknum}.
Equations~\eqref{Matrix3} are particularly interesting, as they show that 
the matrix $M(u)$ can be written as a \textsl{total differential} of simple 
logarithms plus three new functions which are given by
products of complete elliptic integrals and a polynomial in $u$. 
Once appropriate boundary values are known, the integration of the 
system~\eqref{eq:systosolve} as a Laurent series in $(d-4)$ becomes, at least 
in principle, straightforward in terms of iterated integrals with the 
kernels given by the entries of $M(u)$~\eqref{Matrix3}.
Given this result, it is indeed very tempting to try and define a new 
generalized {\it alphabet} composed by the six generalized letters 
appearing in Eq.\eqref{Matrix3}. We will resist the temptation for now, 
and instead go ahead and see how far we can get with what we have.
\par 
We will work for simplicity in the region $0<u<1$ and use everywhere the 
solutions of the homogeneous system valid in this region, $G^{(0,1)}(u)$, 
see appendix~\ref{App:AnCont}.
Working in $0<u<1$ is also very convenient since we can easily fix the 
boundary conditions imposing the regularity of the two original master 
integrals, $h_6(d;u)$ and $h_7(d;u)$, at $u=0^+$ and $u=1^-$. From 
Eqs.~\eqref{eq:sysd4} we can read off the two conditions
\begin{align}
\lim_{u\to 0^+} \Big( h_7(d;u) -\,h_6(d;u) \Big) = 0 \,,\qquad 
\lim_{u \to 1^-} \Big( h_7(d;u) - \frac{16}{3}\,   h_6(d;u) \Big) = 0\,. 
\labbel{eq:boundary}
\end{align}

Having determined~\eqref{eq:boundary}, we can now proceed with the integration
of the differential equations. 
We start from~\eqref{eq:systosolve} and expand everything in $(d-4)$ as follows
\begin{align} 
&m_6(d;u) = m_6^{(0)}(u) + (d-4)\, m_6^{(1)}(u) 
+ \mathcal{O}\left((d-4)^2\right)
\nonumber \\
&m_7(d;u) = m_7^{(0)}(u) + (d-4)\, m_7^{(1)}(u) 
+ \mathcal{O}\left((d-4)^2\right)\,,
\end{align}
such that at order zero the equations reduce to
\begin{align}
\frac{d}{du} \left( \begin{array}{c} m_6^{(0)}(u)\\ m_7^{(0)}(u) 
                    \end{array}\right)
= \frac{1}{\pi} \left( \begin{array}{c} -J_1^\zu(u) \\ 
            I_1^\zu(u) \end{array} \right)\,, \labbel{eq:deqord0}
\end{align} 
and at first order we have instead
\begin{align}
\frac{d}{du} \left( \begin{array}{c} m_6^{(1)}(u)\\ m_7^{(1)}(u) 
                    \end{array}\right)
= \frac{1}{\pi} M(u) \left( \begin{array}{c} m_6^{(0)}(u)\\ m_7^{(0)}(u) 
                     \end{array}\right)\,,  \labbel{eq:deqord1}
\end{align}
where the previous order appears as inhomogeneous term. Note that this 
structure remains true at every order $n$, with $n\geq1$
\begin{align}
\frac{d}{du} \left( \begin{array}{c} m_6^{(n)}(u)\\ m_7^{(n)}(u) 
                    \end{array}\right)
= \frac{1}{\pi} M(u) \left( \begin{array}{c} m_6^{(n-1)}(u)\\ 
                   m_7^{(n-1)}(u)   \end{array}\right)\,, 
\qquad \forall \; n \geq 1\,. \labbel{eq:deqordn}
\end{align}
In the next two sections we describe the integration of~\eqref{eq:deqord0} 
and~\eqref{eq:deqord1}, which will allow us to write a compact result for the
master integrals of the two-loop massive sunrise graph, $h_6(d;u)$ and 
$h_7(d;u)$ up to first order in $(d-4)$. 

\subsection{The two-loop massive sunrise at order zero}
The integration of the order zero, Eq.\eqref{eq:deqord0}, can be carried 
out simply by quadrature. Specializing formulas above in the region $0<u<1$
we find
\begin{align}
m_6^{(0)}(u) = c_6^{(0)}\, -\frac{1}{\pi}\int_0^u\, dt\, J_1^\zu(t)\ , \qquad
m_7^{(0)}(u) = c_7^{(0)}\, +\frac{1}{\pi}\int_0^u\, dt\, I_1^\zu(t)\ . 
\labbel{eq:solmord0}
\end{align}
The constants $c_6^{(0)}$ and $c_7^{(0)}$ can be fixed 
imposing~\eqref{eq:boundary}. Note that the latter must be imposed 
on the original master integrals $h_6(d;u)$ and $h_7(d;u)$ and not on 
$m_6(d;u)$ and $m_7(d;u)$, with the relation between the two sets of 
functions given by Eq.(\ref{eq:rotation}). Expanding also the original 
masters integrals as
\begin{align}
&h_6(d;u) = h_6^{(0)}(u) + (d-4)\, h_6^{(1)}(u) 
+ \mathcal{O}\left((d-4)^2\right)\,,
\nonumber \\
&h_7(d;u) = h_7^{(0)}(u) + (d-4)\, h_7^{(1)}(u) 
+ \mathcal{O}\left((d-4)^2\right)\,,
\end{align}
we find
\begin{align}
h_6^{(0)}(u) &= \frac{1}{\pi}\, 
\left[ J_1^\zu(u) \left( \int_0^u dt\, I_1^\zu(t) + c_7^{(0)} \right) 
     - I_1^\zu(u) \left( \int_0^u dt\, J_1^\zu(t) - c_6^{(0)} \right) 
                                                 \right] \,,\nonumber \\
h_7^{(0)}(u) &= \frac{1}{\pi}\, 
\left[ J_2^\zu(u) \left( \int_0^u dt\, I_1^\zu(t) + c_7^{(0)} \right) 
     - I_2^\zu(u) \left( \int_0^u dt\, J_1^\zu(t) - c_6^{(0)} \right) 
                                                 \right] \,. 
\end{align}
Using the limiting values given in appendix~\ref{App:AnCont} and the definite 
integrals of appendix~\ref{App:DefInt} we obtain
$$c_7^{(0)} = 0\,, \qquad c_6^{(0)} = \int_0^1\,dt\, J^\zu_1(t) = \Cla\,,$$
where the last integral can be performed by standard techniques 
using the integral representation for 
$J_1^\zu(u)$, see for example~\cite{Laporta:2004rb}.
We recall here the definition of the Clausen function 
\begin{equation}
{\rm Cl}_2(x) = - \int_0^x\,  \ln{\left| 2 \sin{\frac{y}{2}}\right|}\, dy
= \frac{i}{2}\left( \Li_2(e^{-i \,x}) - \Li_2(e^{i\,x}) \right)\,.
\end{equation} 
Finally putting everything together we get
\begin{align}
h_6^{(0)}(u) &= \frac{1}{\pi}\, 
\left[ J_1^\zu(u) \int_0^u dt\, I_1^\zu(t) 
     - I_1^\zu(u) \left( \int_0^u dt\, J_1^\zu(t) - \Cla\, \right) 
                                              \right]\,, \nonumber \\
h_7^{(0)}(u) &= \frac{1}{\pi}\, 
\left[ J_2^\zu(u)\int_0^u dt\, I_1^\zu(t)
     - I_2^\zu(u)  \left( \int_0^u dt\, J_1^\zu(t) - \Cla\, \right) 
                                              \right] \,. \labbel{eq:solsun0}
\end{align}
For convenience, we provide here the limiting values of the master
integrals in the two matching points $ u=0^+ $ and $ u=1^- $ 
\begin{align}
&\lim_{u \to 0^+} h_7^{(0)}(u ) = \lim_{u \to 0^+}  h_6^{(0)}(u) 
                                = \frac{1}{\sqrt{3}}\,\Cla\,,
\end{align}
\begin{align}
\lim_{u \to 1^-} h_7^{(0)}(u) = \frac{16}{3}\left( \lim_{u \to 1^-} 
          h_6^{(0)}(u) \right)= \frac{\pi^2}{3}\,.
\end{align} 

The solution~\eqref{eq:solsun0} is valid for $0<u<1$. We can use the 
matching matrices defined in appendix~\ref{App:AnCont} to continue the 
solution in any other region. In particular, it is interesting to study the 
continuation above threshold, i.e for $u>9$ where the master integrals 
develop an imaginary part. By straightforward use of the formulas 
in the appendix we find, for $9<u<\infty$,
\begin{align}
h_6^{(0)}(u) &= \pi J_1^\noo(u) +  \frac{1}{\pi}\, 
\left[ J_1^\noo(u)  \int_9^u dt\, I_1^\noo(t) 
      - I_1^\noo(u)  \left( \int_9^u dt\, J_1^\noo(t) + 5\, \Cla\, 
                                         \right) \right] \nonumber \\
     &+ i\, \pi\, I_1^\noo(u)\,, \nonumber \\& \nonumber \\
h_7^{(0)}(u) &= \pi J_2^\noo(u) +  \frac{1}{\pi}\, 
\left[ J_2^\noo(u)  \int_9^u dt\, I_1^\noo(t) 
      - I_2^\noo(u)  \left( \int_9^u dt\, J_1^\noo(t) + 5\, \Cla\, 
                                         \right) \right] \nonumber \\
     &+ i\, \pi\, I_2^\noo(u)\,,   \labbel{eq:solsun0ph}
\end{align}
such that, as expected, the imaginary parts of the two master integrals at 
order zero in $(d-4)$ read
\begin{align}
\Ima(h_6^{(0)}(u)) = \theta(u-9)\, \pi\, I_1^\noo(u) \,, \qquad
\Ima(h_7^{(0)}(u)) = \theta(u-9)\, \pi\, I_2^\noo(u) \,. \labbel{eq:Imaord0}
\end{align}
Note that the simplicity of the imaginary part above threshold, $u>9$, 
and the absence of an imaginary part for the intermediate region $1<u<9$ 
is true only for the physical masters integrals $h_6(d;u)$ and $h_7(d;u)$. The 
rotated functions, $m_6(d;u)$ and $m_7(d;u)$, have no direct physical meaning
and cannot be expected in general to develop an imaginary part only 
above the $ u>9 $ threshold. 

Having the imaginary part, we can write an alternative representation 
for the solution~\eqref{eq:solsun0} as a dispersion relation
\begin{align}
h_6^{(0)}(u) &= \int_9^\infty\, \frac{dt}{t-u-i\, \epsilon}\, 
                                             I_1^\noo(t)\,,\nonumber \\
h_7^{(0)}(u) &= \frac{1}{\sqrt{3}} \,\Cla + u \left( \frac{5}{6} 
       + \sqrt{3}\,\Cla \right) \nonumber \\ 
  &+  u^2\, \int_9^\infty\, \frac{dt}{t^2(t-u-i\, \epsilon)}\, I_2^\noo(t)\,, 
\labbel{eq:solsun0disp}
\end{align}
where for $h_7^{(0)}(u)$ we have used a doubly subtracted dispersion relation
and fixed the boundary terms matching~\eqref{eq:solsun0disp} 
to~\eqref{eq:solsun0} for $u=0^+$ and $u=1^-$. 
As showed in section~\ref{sec:1lT}, this representation is also particularly 
convenient if we need to integrate once more over it, for example 
whenever the sunrise appears as subtopology in the differential equations of 
more complicated graphs, see section~\ref{sec:kite}.

\subsection{The two-loop massive sunrise at order one}
The order zero of the sunrise graph is special since its inhomogeneous term is
very simple. In order to understand the general structure we want to integrate
Eq.\eqref{eq:deqord1}, which implies integrating over the matrix $M(u)$ 
in~\eqref{Matrix}, using~\eqref{eq:solmord0} as inhomogenous term. Again 
specializing the formulas for $0<u<1$ and integrating by quadrature we get 
\begin{align}
m_6^{(1)}(u) &= \frac{1}{\pi^2} \int_0^u \, dt\, \left[
M_{11}(t)\, \left( \pi\, \Cla -  \int_0^t \, dv\, J_1^\zu(v) \right) 
          + M_{12}(t)  \int_0^t \, dv\, I_1^\zu(v) \right]\,,\nonumber \\
m_7^{(1)}(u) &= \frac{1}{\pi^2} \int_0^u \, dt\, \left[
M_{21}(t)\, \left( \pi\, \Cla -  \int_0^t \, dv\, J_1^\zu(v) \right) 
          + M_{22}(t)  \int_0^t \, dv\, I_1^\zu(v) \right]\,.
\labbel{eq:solmord1a}
\end{align}
At this point the solution is written as a double integral over known 
functions. The entries of the matrix $M(u)$ are rather complicated, 
see~\eqref{Matrix}. Nevertheless the result can be greatly simplified using 
integration by parts identities and the condition on the 
Wronskian~\eqref{eq:Wronsknum}. 
By direct inspection of the matrix~\eqref{Matrix}
it is clear that, at any order $(d-4)^n$, the result can only contain at most 
the following integrals
\begin{align}
&\int_0^u\,dt\, \left\{ t\,,\, 1\,, \frac{1}{t}\,,\, \frac{1}{t-1}\,,\, 
                \frac{1}{t-9} \right\}\, I_1^\zu(t)\,I_1^\zu(t)\, F(t)\,,
\nonumber \\
&\int_0^u\,dt\, \left\{ t\,,\, 1\,, \frac{1}{t}\,,\, \frac{1}{t-1}\,,\, 
                \frac{1}{t-9} \right\}\, I_1^\zu(t)\,J_1^\zu(t)\, F(t)\,,
\nonumber \\
&\int_0^u\,dt\, \left\{ t\,,\, 1\,, \frac{1}{t}\,,\, \frac{1}{t-1}\,,\, 
                \frac{1}{t-9} \right\}\, J_1^\zu(t)\,J_1^\zu(t)\, F(t)\,,
\nonumber \\
&\int_0^u\,dt\,\left\{ \frac{1}{t}\,,\, \frac{1}{t-1}\,,\, \frac{1}{t-9} 
                \right\}\, I_1^\zu(t)\,J_2^\zu(t)\, F(t)\,,
\nonumber \\
&\int_0^u\,dt\,\left\{ \frac{1}{t}\,,\, \frac{1}{t-1}\,,\, \frac{1}{t-9} 
                \right\}\, I_2^\zu(t)\,J_1^\zu(t)\, F(t)\,, \labbel{eq:intsun}
\end{align}
where $F(t)$ is a generic function of $t$ and, at order $n$, it contains the 
order $(n-1)$ of the Laurent expansion of the functions $m_6(d;u)$ and 
$m_7(d;u)$. For $n=1$, which is the case we are interested in, inspection
of Eqs.~\eqref{eq:solmord1a} shows that $F(t)$ is either a constant, or it 
can be one of the two functions
\begin{align}
F(t) =\left\{\, \int_0^t \, dv\, I_1^\zu(v)\,,\, \int_0^t \, dv\, 
             J_1^\zu(v)\, \right\}\,.
\end{align}
Using integration by parts identities together with the condition on the 
Wronskian~\eqref{eq:Wronsknum}, one can show that not all 
integrals~\eqref{eq:intsun} are independent\footnote{Of course, one also
needs to make use of the differential equations satisfied by the 
$I_k(t)$ and $J_k(t)$ in order to re-express the derivatives $d I_k(t)/dt$ 
and $d J_k(t)/dt$ in terms of the $I_k(t)$ and $J_k(t)$.}. 
In particular one can re-express all integrals containing the products 
$I_1(t)\,J_2(t)$ and $I_2(t)J_1(t)$ with the rational prefactors appearing 
in~\eqref{eq:intsun}, in terms of the remaining integrals with 
$I_1(t)I_1(t)$, $I_1(t)J_1(t)$ and $J_1(t)J_1(t)$ only.
This allows to substantially simplify the resulting expressions and, 
notably, eliminate all occurrences of double integrals over the products 
of functions $I_k(t)$ and $J_k(t)$, at the price of introducing simple 
logarithms -- a non trivial result. 
For simplicity we provide here the analytical expressions for the physical 
master integrals only, i.e. $h_6^{(1)}(u)$ and $h_7^{(1)}(u)$, omitting the 
intermediate ones for $m_6^{(1)}(u)$ and $m_7^{(1)}(u)$. The latter can 
anyway easily be recovered by rotating the solution back through the matrix 
$G^{-1}(u)$, see Eq.\eqref{eq:rotation}. \par 
Again in the region $0<u<1$ we can easily fix the boundary values using the 
results of appendices~\ref{App:AnCont} and~\ref{App:DefInt} and we 
find\footnote{As discussed in appendix~\ref{App:DefInt}, we do not 
present explicitly all integrals required to fix all limits. The complete 
list of definite integrals can be obtained by the authors.}

\begin{align}
h_6^{(1)}(u) &= 
\frac{1}{4\,\pi}\,l(u)\,\left( J_1^\zu(u)\, \int_0^u\,dt\, I_1^\zu(t) 
             - I_1^\zu(u)\, \int_0^u\,dt\, J_1^\zu(t) \right) \nonumber \\
&- \frac{1}{4\,\pi}\,\,\left( J_1^\zu(u)\, \int_0^u\,dt\, I_1^\zu(t)\,l(t) 
       - I_1^\zu(u)\, \int_0^u\,dt\, J_1^\zu(t)\,l(t) \right) \nonumber \\
&-\frac{1}{24\,\pi} \left[ \pi^3 - 6\, \Cla\, l(u) + 18\, \Lst \right]\, 
       I_1^\zu(u) - \frac{1}{2}\Cla\, J_1^\zu(u)\,, \labbel{eq:solsun1a}
\end{align}
\begin{align}
h_7^{(1)}(u) &= 
\frac{1}{4\,\pi}\,l(u)\,\left( J_2^\zu(u)\, \int_0^u\,dt\, I_1^\zu(t) 
   - I_2^\zu(u)\, \int_0^u\,dt\, J_1^\zu(t) \right) \nonumber \\
&- \frac{1}{4\,\pi}\,\,\left( J_2^\zu(u)\, \int_0^u\,dt\, I_1^\zu(t)\,l(t) 
  - I_2^\zu(u)\, \int_0^u\,dt\, J_1^\zu(t)\,l(t) \right) \nonumber \\
&-\frac{1}{24\,\pi} \left[ \pi^3 - 6\, \Cla\, l(u) + 18\, \Lst \right] 
  \,I_2^\zu(u) - \frac{1}{2}\Cla\, J_2^\zu(u)\nonumber \\
&+\frac{1}{6\,\pi}\,(u+3)^2\,\left( J_1^\zu(u)\, \int_0^u\,dt\, I_1^\zu(t) 
  - I_1^\zu(u)\, \int_0^u\,dt\, J_1^\zu(t) \right) \nonumber \\
&+ \frac{1}{6\,\pi}\,(u+3)^2\,I_1^\zu(u) \,\Cla, \labbel{eq:solsun1b}
\end{align}
where we introduced the combination of simple logarithms
\be
l(u) = 2\ln{(1-u)} + 2\ln{(9-u)} - \ln{(u)}\,,  \labbel{eq:comblogs}
\ee
and the generalization of the Clausen function
\be
{\rm Ls}_{n}(\theta) = -\int_0^\theta \,dy \,\left[ 
 \ln{ \left( 2\,\sin{\left( \frac{y}{2} \right)} \right)} \right]^{n-1}\,.
\ee
Note that the appearance of the combination $l(u)$ could be foreseen from 
the structure of the matrix $M(u)$ as total differential~\eqref{Matrix3}.

As for the zeroth order, we provide here the boundary values of the two
masters in $u=0^+$ and $u=1^-$
\begin{align}
&\lim_{u \to 0^+} h_7^{(1)}(u ) = \lim_{u \to 0^+}  h_6^{(1)}(u) = 
\sqrt{3} \, \left(  \frac{1}{6}\Cla \ln{(3)} - \frac{1}{4}\,\Lst 
         - \frac{\pi^3}{72} \right)\,,
\end{align}
\begin{align}
\lim_{u \to 1^-} h_7^{(1)}(u) = 
   \frac{16}{3}\left( \lim_{u \to 1^-} h_6^{(1)}(u) \right)=
   \pi^2\, \ln{(2)} - \frac{7}{2}\,\zeta_3\,.
\end{align}

Similarly to the solution at order zero, we can continue 
formulas~\eqref{eq:solsun1a} and~\eqref{eq:solsun1b} above threshold, for 
$u>9$, in order to extract their imaginary parts and use them to write an 
alternative representation of the solutions as dispersion relations. 
Also in this case, the analytic continuation is straightforward using the
results in appendix~\ref{App:AnCont} and for simplicity we give only the
result for the imaginary parts 
\begin{align}
\frac{1}{\pi} \Ima\left( h_6^{(1)}(u)\right) &= \theta(u-9) \left[ 
 \frac{1}{4} \, I_1^\noo(u) \, \bar{l}(u) - \frac{\pi}{2} J_1^\noo(u)\right]
\nonumber \\
\frac{1}{\pi} \Ima\left( h_7^{(1)}(u)\right) &= \theta(u-9) \left[ 
 \frac{1}{4} \, I_2^\noo(u) \, \bar{l}(u) - \frac{\pi}{2} J_2^\noo(u)
+\frac{( u + 3 )^2}{6} \, I_1^\noo(u) \right]\,, \labbel{eq:Imaord1}
\end{align}
where $\bar{l}(u)$ is the real part of the function $l(u)$ defined above 
threshold, i.e. for $u > 9$,
\be
\bar{l}(u) = 2\ln{(u-1)} + 2\ln{(u-9)} - \ln{(u)} \,. \labbel{eq:comblogsb}
\ee
Note that formulas~\eqref{eq:Imaord1} are extremely simple and do not involve any
integral over the functions $I_k(t)$ and $J_k(t)$. 
They allow us to write 
equally simple dispersion relations for the two master integrals
\begin{align}
h_6^{(1)}(u) &= \,\int_9^\infty \frac{dt}{t-u-i\,\epsilon} 
   \left( \frac{1}{4}\,I_1^\noo(t) \, \bar{l}(t)
 - \frac{\pi}{2} \,J_1^\noo(t) \right) \labbel{eq:solsun1dispa} 
\end{align} 
and 
\begin{align} 
h_7^{(1)}(u) &= \sqrt{3}\left[ \frac{1}{6} \Cla \ln{(3)} 
  - \frac{1}{4}\Lst - \frac{\pi^3}{72} \right] \nonumber \\
&+ u\, \left[-\frac{5}{12} + \sqrt{3}\left( \frac{1}{2} \Cla\ln{(3)} 
  - \frac{3}{4}\Lst +\frac{14}{27}\Cla- \frac{\pi^3}{24} \right)\right] 
  \nonumber \\
&+ u^2\, \int_9^\infty \frac{dt}{t^2(t-u-i\,\epsilon)} 
   \left( \frac{1}{4}\,I_2^\noo(t) \, \bar{l}(t)
  - \frac{\pi}{2} \,J_2^\noo(t)  + \frac{(t+3)^2}{6} I_1^\noo(t) \right)\,, 
  \labbel{eq:solsun1dispb}
\end{align}
where, again, the dispersion relation for $h_7^{(1)}(u)$ is 
doubly subtracted in $u=0$.

It is clear that, at least in principle, the techniques described here for 
the integration of the first two orders of the two-loop massive sunrise, 
can be used also for the higher orders.
The formulas are of course more cumbersome and, in general, it is not granted
that the result can always be written in terms of one-fold integrals only, 
as for order zero and one, like in 
Eqs.~\eqref{eq:solsun0},~\eqref{eq:solsun1a} and~\eqref{eq:solsun1b}. 
Nevertheless one can show that, by using integration by parts as we did for 
the order one, also the order $(d-4)^2$ can be substantially simplified.
 
One last comment is in order. The basis of master integrals that we have 
been considering, $h_6(d;u)$ and $h_7(d;u)$, was build by the shift 
$d \to d-2$ of the previous basis $g_6(d;u)$, $g_7(d;u)$, see 
Eq.\eqref{eq:transf2}. That implies that if we expand the latter as Laurent 
series in $(d-2)$
\begin{align}
g_6(d;u) = \sum_{a=0}^\infty\, g_6^{(a)}(u) (d-2)^a\,,\qquad
g_7(d;u) = \sum_{a=0}^\infty\, g_7^{(a)}(u) (d-2)^a\,,
\end{align}
the coefficients of this expansion can be directly related to the 
coefficients of the Laurent expansion in $(d-4)$ of $h_6(d;u)$
and $h_7(d;u)$ as follows
\begin{align}
g_6^{(a)}(u) = h_6^{(a)}(u)\,,\qquad g_7^{(a)}(u) = h_7^{(a)}(u)\,, 
               \qquad \forall\; a\,.
\end{align}
 
\section{The solution for the kite integral} 
\labbel{sec:kite} \setcounter{equation}{0} 
\numberwithin{equation}{section} 
As a last step we will use the results of the previous sections in order to 
write compact expressions for the first two non-zero orders of the kite 
integral. We will do this using the method sketched in section~\ref{sec:1lT}, 
namely we will derive the differential equations for the kite integral and 
then we will insert into it the solution for the sunrise graph given as a
dispersive relation, see Eqs.~\eqref{eq:solsun0disp},~\eqref{eq:solsun1dispa}
and~\eqref{eq:solsun1dispb}.
We start by writing the differential equations for the master integral 
$f_8(d;u)$, defined in~\eqref{eq:basis1}, while for the sunrise we use the 
modified basis defined in~\eqref{eq:transf3}. The differential equation reads
\begin{align}                          
\frac{d}{du}f_8(d;u) &= 
              (d-4)\,\left(\frac{1}{u-1} - \frac{1}{2\,u}\right) f_8(d;u) 
+ \frac{(d-4)^3}{24}\left( 1 - \frac{8}{u-1} \right)\, h_6(d;u) \nonumber \\
&+ \frac{(d-4)}{u-1}\left( - \frac{1}{8}f_1(d;u) + 2\, f_3(d;u) + f_4(d;u) 
            \right)   + (d-4)\frac{1}{u}\, f_5(d;u)\,. \labbel{eq:deqkite1}
\end{align}
Two properties are worth noticing in Eq.\eqref{eq:deqkite1}. First, only 
one of the two master integrals of the sunrise subgraph appears, namely 
$h_6(d;u)$. Second, it appears multiplied by a factor $(d-4)^3$.
This fact is a consequence of the normalization adopted 
in~\eqref{eq:basis1}, where, attempting to build up a basis similar to a 
canonical one, we rescaled all master integrals of suitable powers
of $(d-4)$. 
Note however that, even if in~\eqref{eq:basis1} also $f_6(d;u)$ and $f_7(d;u)$
are rescaled by $(d-4)^2$, one should recall that the masters integrals 
that we are effectively calculating for the sunrise graph (and which enter 
in the differential equation for the kite) are not $f_6(d;u)$ and $f_7(d;u)$, 
but instead $h_6(d;u)$ and $h_7(d;u)$, as defined in~\eqref{eq:transf3}. 
The latter are obtained shifting~\eqref{eq:basis1} from $d \to d-2$, such that
the factor $(d-4)^2$ in front of $f_6(d;u)$ and $f_7(d;u)$ becomes 
effectively a $(d-6)^2$.
In order to make the equations more symmetric, we could have therefore 
rescaled also $h_6(d;u)$ and $h_7(d;u)$ by $(d-4)^2$, reabsorbing in this 
way the corresponding factor in~\eqref{eq:deqkite1}.
We preferred, nevertheless, not to do that in order to avoid the confusion 
of one more change of basis. With the present normalization, the sunrise 
integrals start at order zero in $(d-4)$, which shows that their first 
contribution to the Laurent expansion of the kite integral is at order 
$(d-4)^3$. 
 
We can now move to the actual integration of the equations.
Once more we work in the region $0<u<1$, where the boundary condition can 
be read off directly from Eq.\eqref{eq:deqkite1}, imposing regularity of 
$f_8(d;u)$ on the pseudo-threshold $u=0$. This condition implies
\begin{equation} 
\lim_{u \to 0} f_8(d;u) = 0\,. \labbel{eq:boundkite}
\end{equation}
It is easy to see that $f_8(d;u)$ is finite in $d \to 4$ and therefore its 
Laurent expansion reads
\begin{align}
f_8(d;u) = \sum_{a=0}^\infty f_8^{(a)}(u)\, (d-4)^a \,.
\end{align}
Let us start by looking at the first three orders.
Expanding consistently Eq.\eqref{eq:deqkite1} and inserting the values of the
subtopologies~\eqref{eq:subeasy} we find that for the first three orders
all subtopologies cancel out and we are left with the three chained 
differential equations
\begin{align}
\frac{d}{du}f_8^{(0)}(u) = 0\,,\qquad
 \frac{d}{du}f_8^{(1)}(u) = \left(\frac{1}{u-1} - \frac{1}{2\,u}\right) 
                                        f_8^{(0)}(u)\,, \qquad
 \frac{d}{du}f_8^{(2)}(u) = \left(\frac{1}{u-1} - \frac{1}{2\,u}\right) 
                        f_8^{(1)}(u)
\end{align}
which, together with the boundary condition~\eqref{eq:boundkite}, imply
\begin{align}
f_8^{(0)}(u) = 0\,, \qquad f_8^{(1)}(u) = 0\,, \qquad f_8^{(2)}(u) = 0\,.
\end{align}

The first interesting thing happens at order $(d-4)^3$. Here substituting 
the amplitudes of all the subtopologies except the sunrise integral we are 
left with
\begin{align}
\frac{d}{du}f_8^{(3)}(u) = \frac{1}{24} \left( 1 - \frac{8}{u-1}\right) 
     h_6^{(0)}(u)
+ \frac{1}{u-1}\left( \frac{\pi^2}{96} - \frac{1}{16}G(0,1,u) \right) 
                  + \frac{1}{8\,u} G(1,1,u)\,.
\end{align}
At this point one could, in principle, plug in the solution for the sunrise 
integral as given by~\eqref{eq:solsun0ph}. 
That introduces anyway unneeded complications. The easiest way to proceed 
is instead to insert the dispersive solution, Eq.\eqref{eq:solsun0disp}. 
Upon doing this, the integration in $u$ becomes straightforward in terms 
of multiple polylogarithms and, after fixing the boundary condition, we are 
left (somewhat surprisingly!) with an extremely compact result
\begin{align}
f_8^{(3)}(u) &= \frac{1}{8} G(0,1,1,u) - \frac{1}{16} G(1,0,1,u) 
              - \frac{\pi^2}{96} G(1,u) \nonumber \\
&- \frac{1}{24}\int_9^\infty\, dt\, I_1^\noo(t) \, \left( 1- \frac{8}{t-1} 
      \right)\, G(t,u)\,. \labbel{eq:solkiteord3}
\end{align} 
The very same exercise can be repeated for the next order, making use of the 
dispersion relations derived for the sunrise graph at order 
one~\eqref{eq:solsun1dispa}, and of the previous order just 
computed~\eqref{eq:solkiteord3}.
By integrating the differential equation and fixing the boundary condition 
we get 
\begin{align}
f_8^{(4)}(u) &= \frac{\pi^2}{192} \left( G(0,1,u) - 2 G(1,1,u) \right) + 
\left( \frac{\zeta_3}{32}  + \frac{\pi}{12} \Cla \right) G(1,u)
 - \frac{3}{16} G(0,0,1,1,u) \nonumber \\
&- \frac{1}{32} G(0,1,0,1,u) + \frac{3}{8} G(0,1,1,1,u) 
+ \frac{1}{32}  G(1,0,0,1,u) - \frac{1}{16} G(1,1,0,1,u)  \nonumber \\
&-\frac{1}{96} G(1,u)\, \int_9^\infty\, dt\, I_1^\noo(t) \,\bar{l}(t)
+ \frac{\pi}{48}\, \int_9^\infty\, dt\, J_1^\noo(t) \, 
                          \left( 1- \frac{8}{t-1} \right)\, G(t,u)
\nonumber \\
&-  \frac{1}{96}  \int_9^\infty\, dt\, I_1^\noo(t) \, 
     \left( 1- \frac{8}{t-1} \right) \, \bar{l}(t) \, (G(t,u)- G(1,u)) 
\nonumber \\
&+ \frac{1}{48}   \int_9^\infty\, dt\, I_1^\noo(t) \, 
     \left( 1- \frac{8}{t-1} \right) (G(0,t,u)- 2\, G(1,t,u)) \,, 
\labbel{eq:solkiteord4}
\end{align}
where $\bar{l}(t)$ is defined in~\eqref{eq:comblogsb}.
Note that Eq~\eqref{eq:solkiteord4} contains a combination of, on one side, 
polylogarithms of weight 4 and, on the other, of integrals over elliptic 
integrals and polylogarithms of weight 2.

\subsection{The analytic continuation of the solution}
Here we want to show that also in this case the analytic continuation of our 
solution, Eqs.~\eqref{eq:solkiteord3} and~\eqref{eq:solkiteord4}, is 
completely straightforward in the whole range $-\infty < u < +\infty$. The 
kite integral has a first cut at $u=1$ corresponding to $s = m^2$, where the 
harmonic polylogarithms develop an imaginary part.
The second cut is at $u=9$, $s=9\, m^2$, and the elliptic integrals develop 
further imaginary parts which can be easily computed using the results of 
appendix~\ref{App:AnCont}.
Let us consider for example the first non-zero order, 
Eq.\eqref{eq:solkiteord3}, and let us continue it in the two physically 
relevant regions, i.e. for $1<u<9$ and then above the three-mass 
threshold $9<u<\infty$.

\begin{itemize}
\item[a)] The region $1<u<9$.

In this region the HPLs develop an imaginary part, whose sign is fixed by 
Feynman's prescription $u \to u + i 0^+$. On the other hand, the pieces 
containing the integration over the imaginary part of the sunrise remain 
real since $G(t,u) \in \mathbb{R}$ if $t>u$. 
In order to obtain real-valued polylogarithms it is convenient to perform 
the change of variables
\be
v = \frac{u-1}{8}\,, \qquad \mbox{such that} 
         \qquad 1<u<9\quad \to \quad 0<v<1\,.
\ee

The analytic continuation of the HPLs then gives
\begin{align}
f_8^{(3)}(u)\Big|_{1<u<9} &= \frac{\zeta_3}{4} 
+ \frac{1}{16}\left[ 9\,\ln^2{(2)} - \pi^2 \right]G(-1/8,v) 
\nonumber \\
&+ \frac{3}{8} \ln{(2)} \left[ G(-1/8,0,v) - \frac{1}{2}G(0,-1/8,v) 
                          \right]\nonumber \\
&+ \frac{1}{8}G(-1/8,0,0,v) - \frac{1}{16} G(0,-1/8,0,v) \nonumber \\
&- \frac{1}{24}\int_9^\infty\, dt\, I_1^\noo(t) \, 
               \left( 1- \frac{8}{t-1} \right)\, G(t,u)\,\nonumber \\
&+ i\, \pi \left[ \frac{1}{16} G(0,-1/8,v) - \frac{1}{8} G(-1/8,0,v) 
                 - \frac{3}{8} \ln{(2)}\, G(-1/8,v)\right]\,.
\end{align}

\item[b)] The region $9<u<\infty.$

The analytic continuation to this region involves also the continuation
over the elliptic kernels coming from the sunrise graph. For $u>9$, the 
logarithm $G(t,u)$ develops an imaginary part whenever $t < u$. To keep 
track of this, it is enough to split the corresponding integral in $t$ 
into two pieces 
\begin{align}
\int_9^\infty\, dt\, I_1^\noo(t) \, \left( 1- \frac{8}{t-1} \right)\, G(t,u) 
&=  \int_9^u \, dt\, I_1^\noo(t) \, \left( 1- \frac{8}{t-1} \right)\, 
                  \ln{\left( 1 - \frac{u+i\,0^+}{t} \right)} \nonumber \\
&+  \int_u^\infty dt\, I_1^\noo(t) \, \left( 1- \frac{8}{t-1} \right)\, 
                                                         G(t,u) \nonumber \\
&= \int_9^u \, dt\, I_1^\noo(t) \, \left( 1- \frac{8}{t-1} \right)\, 
                             \ln{\left( \frac{u}{t} -1\right)}  \nonumber \\
&+  \int_u^\infty dt\, I_1^\noo(t) \, \left( 1- \frac{8}{t-1} \right)\, 
                                                          G(t,u) \nonumber\\
&- i\, \pi \, \int_9^u \, dt\, I_1^\noo(t) \, \left( 1- \frac{8}{t-1} 
                                              \right)\,,
\end{align}
where we used, as always, $u \to u + i 0^+$. 
On the other hand, the multiple-polylogarithms of $v$ remain real since, 
for $9<u<\infty$, we have $1<v<\infty$ and all multiple polylogarithms of 
$v$ have only a cut in $v = -1/8$. Putting everything together we find
\begin{align}
f_8^{(3)}(u)\Big|_{9<u<\infty} &= \frac{\zeta_3}{4} 
+ \frac{1}{16}\left[ 9\,\ln^2{(2)} - \pi^2 \right]G(-1/8,v) 
\nonumber \\
&+ \frac{3}{8} \ln{(2)} \left[ G(-1/8,0,v) - \frac{1}{2}G(0,-1/8,v) 
                                                   \right]\nonumber \\
&+ \frac{1}{8}G(-1/8,0,0,v) - \frac{1}{16} G(0,-1/8,0,v)  \nonumber \\
&- \frac{1}{24}\int_9^u\, dt\, I_1^\noo(t) \, \left( 1- \frac{8}{t-1} 
           \right)\,  \ln{\left( \frac{u}{t} -1\right)}\, \nonumber \\
&- \frac{1}{24}\int_u^\infty\, dt\, I_1^\noo(t) \, \left( 1- \frac{8}{t-1} 
                                       \right)\,  G(t,u)\,\nonumber \\ 
&+ i\, \pi \left[ \frac{1}{16} G(0,-1/8,v) - \frac{1}{8} G(-1/8,0,v) 
            - \frac{3}{8} \ln{(2)}\, G(-1/8,v)  \right. \nonumber \\ 
&\left.  \qquad + \frac{1}{24}\int_9^u\, dt\, I_1^\noo(t) \, 
                           \left( 1- \frac{8}{t-1} \right) \right]\,.
\end{align}
\end{itemize}

The very same steps can be repeated in order to obtain the analytic 
continuation of the next order, Eq.\eqref{eq:solkiteord4}. We do not report 
the results here for conciseness.
\section{Conclusions}  \labbel{sec:concl} \setcounter{equation}{0} 
\numberwithin{equation}{section}
The computation of multiloop massive Feynman integrals remains still today an outstanding task
due to the appearance of new mathematical structures which cannot be reduced to the 
by now very well understood multiple polylogarithms. 
The best known example is that of the two-loop massive sunrise graph. In spite of the 
recent impressive progress, a formalism which allows to treat not only the sunrise graph,
but also, more importantly, more complicated diagrams which, for example, contain it as subgraph, 
is still missing in the literature.
This issue, indeed, becomes of crucial importance for LHC phenomenology, whenever 
the contribution of massive particles in the loops has to be taken into account.\par
In this paper we showed that the study of 
the imaginary part of Feynman graph amplitudes, and the corresponding dispersion relations,
can be paired to the differential equations method, providing a very powerful tool for the evaluation of 
massive Feynman integrals, in particular when the result cannot be written in terms of multiple
polylogarithms only. We have considered in detail the case of the kite graph, relevant for the calculation
of the two-loop QED corrections to the electron self-energy. The calculation of the kite integral
within the differential equations method requires the integration over its full set of subgraphs, 
which contain both simple integrals which can be expressed in terms of harmonic polylogarithms,
and the two-loop massive sunrise. While the former do not constitute
any conceptual difficulty and can be treated with standard techniques, the latter
require the extension of these techniques. After having established the formalism
for the solution of the coupled differential equations satisfied by the two master integrals
of the sunrise graph, we showed how to compute their imaginary part and write dispersive
relations for the latter. Finally we used these results in order to obtain simple analytical
representations for the first two non-zero orders of the kite integral.
The final expressions involve polylogarithms up to weight 4 and one-fold integrals over complete elliptic
integrals and polylogarithms of weight 2. The numerical evaluation of our result is 
straightforward, as well as their analytic continuation to all physically relevant values of
the momentum squared.\par
While the problem studied in this paper is relatively simple, the methods presented are very general
and can be, in principle, easily extended to consider arbitrarily complicated cases.
Moreover, the results derived here, in particular the expressions for the two master integrals of the two-loop
massive sunrise, are in a form that is suitable to be re-used once they appear
as inhomogeneous terms in the differential equations of more complicated graphs. 
The application of these techniques to phenomenologically
relevant three- and four-point functions is currently under study.

\section*{Acknowledgements}
We are grateful to J. Vermaseren for his assistance in the use of the algebraic program 
FORM~\cite{Vermaseren:2000nd} which was intensively used in all the steps of the calculation.
All analytical results for the master integrals have been checked numerically with SecDec 3~\cite{Borowka:2015mxa}.
We are grateful to Andreas von Manteuffel and Pierpaolo Mastrolia for discussions at different stages
of the project and for their comments to the manuscript.
One of the authors (E.R.) acknowledges the pleasant stays offered to him by the Erwin Schr\"{o}dinger Institute, Wien,
and the Institut f\"{u}r Theoretische Teilchenphysik of KIT, Karlsruhe, where part of the ideas 
of this paper were developed. 
\appendix

\section{Elliptic integrals}\labbel{App:Ell} \setcounter{equation}{0} 
\numberwithin{equation}{section}
For convenience of the reader, we collect in this Appendix a number of 
results on elliptic integrals, written in the notation that we use 
throughout the paper, following~\cite{Laporta:2004rb} (and fixing some 
misprints occurring there). \par 
Quite in general, consider the fourth-order polynomial
\begin{align}
R_4(b) = (b-b_1)(b-b_2)(b-b_3)(b-b_4), \labbel{A1} 
\end{align}
where the four real constants $ b_i $ satisfy the condition 
$b_1< b_2< b_3< b_4$. 
We can define three apparently different integrals
\begin{align}
J(b_1,b_2,b_3,b_4) = &\int_{b_1}^{b_2} \frac{db}{\sqrt{-R_4(b)}}\,, \quad
I(b_1,b_2,b_3,b_4) = \int_{b_2}^{b_3} \frac{db}{\sqrt{R_4(b)}}\,, \nonumber \\
&K(b_1,b_2,b_3,b_4) = \int_{b_3}^{b_4} \frac{db}{\sqrt{-R_4(b)}}\,, 
\end{align} 
but in fact they are not all independent. Indeed, consider the contour 
integral 
\be C = \oint \frac{db}{\sqrt{R_4(b)}} \ , \labbel{oR4} \ee 
where the contour contains the four points $ b_i $. The integrand has 
two cuts, one cut from $ b_1 $ to $ b_2 $, where 
$ R_4(b+i\eps) = - iR_4(-b) \ ,$ the other cut from $ b_3 $ to $ b_4 $ 
with $ R_4(b+i\eps) = iR_4(-b) \ .$ 
If the contour is the circle at infinity, where $ 1/R_4(b) $ behaves as 
$ 1/b^2 $, one finds 
$$ C = 0 \ . $$ 
By shrinking the circle to two closed paths containing one of the cuts 
each, one obtains 
$$ C = -2iJ(b_1,b_2,b_3,b_4) + 2iK(b_1,b_2,b_3,b_4) \ ; $$ 
by comparing the two results for $ C $, one has in general 
\be 
J(b_1,b_2,b_3,b_4)  = K(b_1,b_2,b_3,b_4)\,. \labbel{eq:dipfun} 
\ee 

In the case of the equal-mass sunrise the polynomial becomes
\begin{align}
R_4(b) = R_4(b,u) = b(b-4)(b-(\sqrt{u}-1)^2)(b-(\sqrt{u}+1)^2) \labbel{A5} 
\end{align}
such that $$R_4(b,u) > 0 \qquad \mbox{if} \qquad 4<b<(\sqrt{u}-1)^2.$$
We define, for $n$ integer and positive, the following three functions
\begin{align}
&J(n,u) = \int_0^4 db\, \frac{b^n}{\sqrt{-R_4(b,u)}}\nonumber\\& \nonumber \\
&I(n,u) = \int_4^{(\sqrt{u}-1)^2} db\, \frac{b^n}{\sqrt{R_4(b,u)}}\nonumber\\& \nonumber \\
&K(n,u) = \int_{(\sqrt{u}-1)^2} ^{(\sqrt{u}+1)^2} 
db\, \frac{b^n}{\sqrt{-R_4(b,u)}} \ , \labbel{A6} 
\end{align} 
such that they are all real-valued as $u>9$. 
Clearly, not all functions are linear independent.
Using integration-by-parts identities
$$\int_{\beta_1}^{\beta_2} db \frac{d}{db}\left(\,b^n\, \sqrt{R_4(b,u)} \right)= 0\,, 
\qquad \forall \beta_i \in \{ 0,\, 4,\, (\sqrt{u}-1)^2,\, (\sqrt{u}+1)^2\}\,,$$
it is easy to prove that, for each family of functions, only three can be linear independent. We
choose for definiteness
\begin{align}
&J(0,u),\,\, J(1,u),\,\, J(2,u)\,,  \nonumber \\ 
&I(0,u),\,\, I(1,u),\,\, I(2,u)\,,  \nonumber \\ 
&K(0,u),\,\, K(1,u),\,\, K(2,u)\,.
\end{align}
Moreover one more relation can be written for each family of functions. We find 
\begin{align}
\int_4^{(\sqrt{u}-1)^2} db \, \frac{d}{db}\,\ln{ \left( \frac{b(u+3-b) + \sqrt{R_4(b,u)} }{b(u+3-b) - \sqrt{R_4(b,u)} }\right) }
= \int_4^{(\sqrt{u}-1)^2} db  \,\frac{(3b-u-3)}{\sqrt{R_4(b,u)}} = 0\,,
\end{align}
\begin{align}
\int_0^4 db \, \frac{d}{db}\,\ln{ \left( \frac{b(u+3-b) +i\, \sqrt{-R_4(b,u)} }{b(u+3-b) - i\,\sqrt{-R_4(b,u)} }\right) }
= \int_0^4 db  \,\frac{i\,(3b-u-3)}{\sqrt{R_4(b,u)}} = -i\,\pi\,,
\end{align}
\begin{align}
\int_{\bb}^{\BB} db \, \frac{d}{db}\,\ln{ \left( \frac{b(u+3-b) +i\, \sqrt{-R_4(b,u)} }{b(u+3-b) - i\,\sqrt{-R_4(b,u)} }\right) }
= \int_{\bb}^{\BB} db  \,\frac{i\,(3b-u-3)}{\sqrt{R_4(b,u)}} = 2\,i\,\pi\,,
\end{align}
which imply respectively
\begin{align}
J(1,u) = \frac{(u+3)}{3} J(0,u) - \frac{\pi}{3}\,,\nonumber
\end{align}
\begin{align}
I(1,u) = \frac{(u+3)}{3} I(0,u)\,,\nonumber
\end{align}
\begin{align}
K(1,u) = \frac{(u+3)}{3} K(0,u) + \frac{2\,\pi}{3}\,.
\end{align}
Finally, as expected from~\eqref{eq:dipfun}, 
one can prove that the functions $K(n,u)$ and $J(n,u)$
are not linearly independent from each other, in particular it holds
\begin{align}
 &K(0,u) = J(0,u)\,,\nonumber \\
 &K(1,u) = J(1,u) + \pi \,,\nonumber \\
 &K(2,u) = J(2,u) + \pi(u+3)\,.
 \end{align}
All together these relations imply that only 4 functions are linearly independent.
We choose our basis as follows
$$I(0,u),\,\, I(2,u),\,\,J(0,u),\,\, J(2,u).\,\,$$

\section{The analytic continuation of the homogeneous solutions} 
\labbel{App:AnCont} \setcounter{equation}{0} 
\numberwithin{equation}{section}
In the main text we showed how to find the solution of the homogeneous 
system for the sunrise graph, the matrix $G^{(9,\infty)}(u)$ in the 
region $9<u<\infty$, Eq.(\ref{defG9oo}), 
using the imaginary part of the master integrals as building blocks. 
In this appendix we show how
to build up corresponding real solutions in the remaining three regions, i.e. $1<u<9$,
$0<u<1$ and $-\infty < u < 0$.

\subsection{The region with $0<u<1$}
In this region the 4 roots of $R_4(b,u)$ are ordered as $\{\,0,(\sqrt{u}-1)^2,4,(\sqrt{u}+1)^2\,\}$. We choose 
therefore as solutions again the ones going between the first two roots, namely

\begin{align}
 &I_1^\zu(u) = 
 \int_{(\sqrt{u}-1)^2}^{(\sqrt{u}+1)^2} \frac{db}{\sqrt{R_4(b,u)}}\nonumber\\
 &I_2^\zu(u) =  
 \int_{(\sqrt{u}-1)^2}^{(\sqrt{u}+1)^2} \frac{db\,b^2}{\sqrt{R_4(b,u)}}\nonumber\\
 &J_1^\zu(u) = \int_0^{(\sqrt{u}-1)^2}         \frac{db}{\sqrt{-R_4(b,u)}}\nonumber \\
 &J_2^\zu(u) =  \int_0^{(\sqrt{u}-1)^2}       \frac{db\,b^2}{\sqrt{-R_4(b,u)}} + \frac{\pi}{3}(u+3)\,.
\end{align}

Let us compute again the limits on the boundaries of the region of definition. 

\subsubsection{Limits for $u \to 1^-$}
As $u \to 1^-$ we find (keeping the leading logarithmic behaviour):

\begin{align}
 &I_1^{\zu}(u\to1^-) = \frac{3}{4}\left( 3 \ln{2} - \ln{(1-u)}\right)\,,\nonumber \\
 &I_2^{\zu}(u\to1^-)  = -4 + 12 \ln{2} - 4 \ln{(1-u)} \nonumber\\
 &J_1^{\zu}(u\to1^-) = \frac{\pi}{4} \,,\qquad
 J_2^{\zu}(u\to1^-) = \frac{4}{3}\pi \,,
\end{align}
which give again for the Wronskian
\begin{equation}
\lim_{u \to 1^-}W^\zu(u )  = \pi. \labbel{B8} 
\end{equation}

\subsubsection{Limits for $u \to 0^+$}
As $u \to 0^+$ we find (keeping the leading logarithmic behaviour):

\begin{align}
 &I_1^{\zu}(u\to0^+)  = \frac{\sqrt{3}}{3}\pi \,,\qquad 
   I_2^{\zu}(u\to0^+)  = \frac{\sqrt{3}}{3}\pi \,,\nonumber\\
 &J_1^{\zu}(u\to0^+) = \sqrt{3}\left( \frac{\ln{3}}{3} - \frac{\ln{u}}{6}\right)\,,\nonumber\\
 &J_2^{\zu}(u\to0^+) = \sqrt{3}\left( \frac{\ln{3}}{3} - \frac{\ln{u}}{6} + 1 \right) \,,
\end{align}
and the Wronskian is again
\begin{equation}
\lim_{u \to 0^+}W^\zu(u)  = \pi. \labbel{B10} 
\end{equation}

\subsection{The region with $1<u<9$}
In this region the 4 roots of $R_4(b,u)$ are ordered as 
$\{\,0,(\sqrt{u}-1)^2,4,(\sqrt{u}+1)^2\,\}$. We choose 
therefore as solutions again the ones going between the first two roots, 
namely

\begin{align}
 &I_1^\un(u) = 
 \int_{(\sqrt{u}-1)^2}^4 \frac{db}{\sqrt{R_4(b,u)}}\nonumber\\
 &I_2^\un(u) =  
 \int_{(\sqrt{u}-1)^2}^4 \frac{db\,b^2}{\sqrt{R_4(b,u)}}\nonumber\\
 &J_1^\un(u) = \int_0^{(\sqrt{u}-1)^2} \frac{db}{\sqrt{-R_4(b,u)}}\nonumber\\
 &J_2^\un(u) = \int_0^{(\sqrt{u}-1)^2} \frac{db\,b^2}{\sqrt{-R_4(b,u)}} 
             + \frac{\pi}{3}(u+3)\,. \labbel{B1} 
\end{align}

Let us compute again the limits on the boundaries of the region of 
definition. 

\subsubsection{Limits for $u \to 9^-$}
As $u \to 9^-$ we find (keeping the leading logarithmic behaviour):

\begin{align}
 &I_1^{\un}(u\to9^-) = \frac{\sqrt{3}}{12}\pi\,,\qquad 
 I_2^{\un}(u\to9^-)  = \frac{4\sqrt{3}}{3}\pi \nonumber\\
 &J_1^{\un}(u\to9^-) =   
 \frac{\sqrt{3}}{2} \left( \frac{\ln{3}}{3} + \frac{\ln{2}}{2} 
                        - \frac{\ln{(9-u)}}{6}\right) \nonumber\\
 &J_2^{\un}(u\to9^-) = 
  4 \sqrt{3} \left( 1 + \frac{2\,\ln{3}}{3} + \ln{2} 
                        - \frac{\ln{(9-u)}}{3}\right) \,,
\end{align}
which gives again for the Wronskian
\begin{equation}
\lim_{u \to 9^-}W^\un(u)  
= \pi.
\end{equation}

\subsubsection{Limits for $u \to 1^+$}
As $u \to 1^+$ we find (keeping the leading logarithmic behaviour):

\begin{align}
 &I_1^{\un}(u\to1^+) = \frac{3}{4}\left( 3 \ln{2} - \ln{(u-1)}\right)\,, 
                                                      \nonumber\\
 &I_2^{\un}(u\to1^+)  = -4 + 12 \ln{2} - 4 \ln{(u-1)} \nonumber\\
 &J_1^{\un}(u\to1^+) = \frac{\pi}{4} \,,\qquad
 J_2^{\un}(u\to1^+) = \frac{4}{3}\pi \,,
\end{align}
which give again for the Wronskian
\begin{equation}
\lim_{u \to 1^+}W^\un(u)  = \pi. \labbel{B5} 
\end{equation}

\subsection{The region with $u=-z<0$}
Last but not least we must consider the non-physical euclidean region, namely $u = -z <0$.
In this region two of the 4 roots become complex, in particular we have
$\{\,0,4\,\}$ and $\{\, (\sqrt{-z}-1)^2 = 1 +z - 2\,i\,\sqrt{z},\, (\sqrt{-z}+1)^2 = 1 +z + 2\,i\,\sqrt{z} \,\}.$
This implies as well that two solutions are one the complex conjugate of the other

\begin{align*}
\left( \int_0^{(\sqrt{u}-1)^2}  \frac{db}{\sqrt{-R_4(b,u)}} \right)^* = 
\int_0^{(\sqrt{u}+1)^2}  \frac{db}{\sqrt{-R_4(b,u)}} \,,
\end{align*}
where $u$ now is negative.
Since both integrals develop imaginary parts in this region, but we know that the final result must be real,
we choose as linear independent solutions the following real combinations

\begin{align*}
&I_1^\ooz(u) = \frac{1}{i}\left( \int_0^{(\sqrt{u}+1)^2}  \frac{db}{\sqrt{-R_4(b,u)}} 
-  \int_0^{(\sqrt{u}-1)^2}  \frac{db}{\sqrt{-R_4(b,u)}} \right)
 \nonumber\\
 &I_2^\ooz(u) =  \frac{1}{i}\left( \int_0^{(\sqrt{u}+1)^2}  \frac{db\,b^2}{\sqrt{-R_4(b,u)}} 
-  \int_0^{(\sqrt{u}-1)^2}  \frac{db\,b^2}{\sqrt{-R_4(b,u)}} \right)
 \nonumber\\
 &J_1^\ooz(u) =\frac{1}{2}\left( \int_0^{(\sqrt{u}+1)^2}  \frac{db}{\sqrt{-R_4(b,u)}} 
+  \int_0^{(\sqrt{u}-1)^2}  \frac{db}{\sqrt{-R_4(b,u)}} \right)
 \nonumber \\
 &J_2^\ooz(u) =   \frac{1}{2}\left( \int_0^{(\sqrt{u}+1)^2}  \frac{db\,b^2}{\sqrt{-R_4(b,u)}} 
+  \int_0^{(\sqrt{u}-1)^2}  \frac{db\,b^2}{\sqrt{-R_4(b,u)}} \right)+ \frac{\pi}{3}(u+3)\,.
\end{align*}

We need once more to study the limits of these four solutions on the boundaries, namely $u \to -\infty$ and
$u \to 0^-$.

\begin{align}
 &I_1^{\ooz}(u\to0^-)  = \frac{\sqrt{3}}{3}\pi \,,\qquad 
   I_2^{\ooz}(u\to0^-)  = \frac{\sqrt{3}}{3}\pi \,,\nonumber\\
 &J_1^{\ooz}(u\to0^-) = \sqrt{3}\left( \frac{\ln{3}}{3} - \frac{\ln{(-u)}}{6}\right)\,,\nonumber\\
 &J_2^{\ooz}(u\to0^-) = \sqrt{3}\left( \frac{\ln{3}}{3} - \frac{\ln{(-u)}}{6} + 1 \right) \,,
\end{align}
which give again for the Wronskian
\begin{equation}
\lim_{u \to 0^-}W^\ooz(u )  
= \pi.
\end{equation}

\begin{align}
 &I_1^{\ooz}(u\to-\infty) = - 3 \frac{\ln{(-u)}}{u}\,,\qquad 
 I_2^{\ooz}(u\to-\infty)  =  -\,u\,\ln{(-u)} + 2\,u \nonumber\\
 &J_1^{\ooz}(u\to-\infty) =   -\frac{\pi}{2\,u}\,,\qquad \qquad
 J_2^{\ooz}(u\to-\infty) = - \frac{\pi}{6} u\,,
 \end{align}
which of course give once more
\begin{equation}
\lim_{u \to -\infty}W^\ooz(u)   = \pi. \labbel{B14} 
\end{equation}

\subsection{Matching}
As a last step we must write down the matrices which allow to match the solutions in
the different regions, and therefore analytically continue them to the whole range $-\infty < u< \infty$.
In order to do this, we assign a positive imaginary part to $u \to u + i 0^+$ throughout the paper.
Let $G^{(a,b)}(u)$ be the $2 \times 2$ matrix of real solutions valid for $a < u < b$

\begin{equation}
G^{(a,b)}(u) = \left( \begin{array}{cc}  I_1^{(a,b)}(u) & J_1^{(a,b)}(u) \\
 I_2^{(a,b)}(u) & J_2^{(a,b)}(u) \end{array} \right)\,,
\end{equation}
and $M^{(b)}$ be the matching matrix in the point $u = b$. We have then that, given a set of solutions
valid in the interval $a<u<b$, these can be continued to the interval $b<u<c$ as
\begin{equation}
G^{(b,c)}(u) = G^{(a,b)}(u)\,M^{(b)}\,, \labbel{eq:matching1}
\end{equation}
where the matching is performed in the point $u=b$ through the matrix $M^{(b)}$.

Using the
limits computed in the previous paragraph and using $u \to u + i 0^+$, we obtain, starting from $u=0$
\begin{align}
&M^{(0)} = \left( 
\begin{array}{cc} 1 & -i/2  \\ 0 & 1 \end{array} 
\right)\,,
\qquad 
M^{(1)} = \left( 
\begin{array}{cc} 1 & 0  \\ -3\,i & 1 \end{array} 
\right) \nonumber \\
&M^{(9)} = \left( 
\begin{array}{cc} 1 & -i  \\ 0 & 1 \end{array} 
\right)\,,\qquad 
M^{(\infty)} = \left( 
\begin{array}{cc} -2 & 0 \\ 3\,i & - 1/2 \end{array} 
\right)\,, \labbel{eq:matching2}
\end{align}
and one finds, of course
\begin{equation}
M^{(0)}\,M^{(1)}\,M^{(9)}\,M^{(\infty)} = \left( 
\begin{array}{cc} 1 & 0\,  \\ 0 & 1 \end{array} 
\right)\,.
\end{equation}

\section{Some definite integrals}\labbel{App:DefInt} \setcounter{equation}{0} 
\numberwithin{equation}{section}
In this section we collect some results on relevant definite integrals over
the functions defined in the previous section.  This list is, necessarily, incomplete.
The complete list of all integrals
necessary for the computations described in the paper can be obtained from the authors.
We stress here that all these integrals
can be computed by suitable application of the methods
described in~\cite{Laporta:2004rb}.

\begin{itemize}
\item Definite integrals for $0<u<1$
\begin{align}
&\int_0^1\, dt\, I_1^{\zu}(t) = \frac{\pi^2}{4}\,, \qquad \int_0^1\, dt\, I_2^{\zu}(t) = \frac{3\,\pi^2}{2} - 8\,,\nonumber \\
&\int_0^1\, dt\, J_1^{\zu}(t) = \Cla\,, \qquad \int_0^1\, dt\, J_2^{\zu}(t) =-\frac{11 \sqrt{3}}{8}
+ 6\, \Cla\,.
\end{align}

\item Definite integrals for $1<u<9$
\begin{align}
&\int_1^9\, dt\, I_1^{\noo}(t) = \frac{3\,\pi^2}{4}\,, \qquad \int_1^9\, dt\, I_2^{\un}(t) = \frac{9\,\pi^2}{2} + 8\,,\nonumber \\
&\int_1^9\, dt\, J_1^{\un}(t) = 5\,\Cla\,, \qquad \int_1^9\, dt\, J_2^{\un}(t) = 28 \sqrt{3}
+ 30 \, \Cla\,.
\end{align}

\item Definite integrals for $9<u<\infty$. Since the integral are divergent, we
introduce a cutoff $U\gg1$ and we find
\begin{align}
&\int_9^U\, dt\, I_1^{\noo}(t) = \frac{3\,\ln{U}^2}{4} - \frac{\pi^2}{4}\,, \nonumber\\
&\int_9^U\, dt\, I_2^{\noo}(t) = \frac{1}{4} U^2 \ln{U}  - \frac{5\,U^2}{8}  + \frac{3}{2} U \ln{U} -\frac{U}{2}  
+ \frac{9}{2} \ln{U}^2 - 7 \ln{U} -\frac{13}{8} - \frac{3\,\pi^2}{2}
 \,,\nonumber \\
&\int_9^U\, dt\, J_1^{\noo}(t) = \pi \ln{U} - 5 \Cla\,,  \nonumber \\
&\int_9^U\, dt\, J_2^{\noo}(t) =  \frac{\pi}{6} U^2 + \pi \, U + 6\,\pi\,\ln{U}  -28\,\sqrt{3} - \frac{14\,\pi}{3} - 30\Cla\,.
\end{align}

\end{itemize}
Similar expressions can be found for integrals containing the functions $I_k^{(a,b)}(u)$ and $J_k^{(a,b)}(u)$
together with the different rational factors $1/u$, $1/(u-1)$, $1/(u-9)$ and with the three corresponding
logarithms $\ln{|u|}$, $\ln{|u-1|}$ and $\ln{|u-9|}$.

\section{Relation with the complete elliptic integrals} \labbel{App:CEllInt} 
\setcounter{equation}{0} 
\numberwithin{equation}{section}
In this last Appendix we show how to express the solutions entering 
in the $G^{(a,b)}(u)$ for all four relevant intervals $a<u<b$, 
see Eq.s(\ref{defG},\ref{defG9oo}), in terms of the complete elliptic 
integrals of first and second kind. The latter are defined as 
\begin{align} 
K(x) &= \int_0^1 \frac{dt}{ \sqrt{(1-t^2)(1- x \,t^2)}}\,, \qquad 
E(x) = \int_0^1 dt \frac{\sqrt{1- x \,t^2}}{ \sqrt{1-t^2}}\,. 
\end{align} 
They are real for $0<x<1$. From the very definition, one has the particular 
values 
\begin{align} K(0) = \frac{\pi}{2} \ , \hspace{2cm} 
             &K(1-\eta) = 2\, \ln{2} - \frac{1}{2}\ln\eta \ , \nonumber\\ 
 E(0) = \frac{\pi}{2} \ , \hspace{2cm} &E(1) = 1 \ , \labbel{valKE} 
\end{align} 
where $ \eta $ is small and positive and terms of first order in $ \eta $ 
are neglected. For $ \eta = -\xi-i\eps $, with $ \xi$ small and positive, 
$ \eps > 0 $ and infinitesimal, the above equation for $ K(1-\eta) $ 
gives further 
\be K(1+\xi+i\eps) =  2\, \ln{2} - \frac{1}{2}\ln\xi + i \, 
\frac{\pi}{2} \ . \labbel{K(1+eta)} \ee 
$ K(x), E(x) $ satisfy the system of first order differential equations 
given by 
\be \frac{d}{dx} \left( \begin{array}{c} K(x) \\ E(x) \end{array} \right) 
   = \frac{1}{2x}\left( \begin{array}{cc} - 1 & \frac{1}{1-x} \\ 
                         -1 & 1 \end{array} \right) 
                 \left( \begin{array}{c} K(x) \\ E(x) \end{array} \right) \ . 
\labbel{KEeqs} \ee 
Considering, more in general, the differential system 
\be \frac{d}{dx} \left( \begin{array}{c} F_1(x) \\ F_2(x) \end{array} \right) 
   = \frac{1}{2x}\left( \begin{array}{cc} - 1 & \frac{1}{1-x} \\ 
                         -1 & 1 \end{array} \right) 
            \left( \begin{array}{c} F_1(x) \\ F_2(x) \end{array} \right) \ , 
\labbel{F12sys} \ee 
the pair of functions $ (K(x), E(x)) $, obviously  provides with a first 
solution, say $ F^{(1)}_i(x), i=1,2 $ 
$$ \left( \begin{array}{c} F^{(1)}_1(x) \\ F^{(1)}_2(x) \end{array} \right) 
 = \left( \begin{array}{c} K(x) \\ E(x) \end{array} \right) \ , $$ 
while a simple calculation shows that 
\begin{align} 
   \left( \begin{array}{c} F^{(2)}_1(x) \\ F^{(2)}_2(x) \end{array} \right) 
&= \left( \begin{array}{c} K(1-x) \\ K(1-x)-E(1-x) \end{array} \right) \ , 
\nonumber\\ 
\left( \begin{array}{c} F^{(3)}_1(x) \\ F^{(3)}_2(x) \end{array} \right) 
&= \frac{1}{\sqrt{x}} \left( \begin{array}{c} K(\frac{1}{x}) \\ 
     (1-x)K(\frac{1}{x}) + xE(\frac{1}{x}) \end{array} \right) \ , 
\nonumber 
\end{align} 
are also solutions. They cannot be all independent, and in fact one has,
for $x \to x + i\,\epsilon$, with $0<x<1$, the 
relation 
\be F^{(3)}_i(x) = F^{(1)}_i(x) - iF^{(2)}_i(x) \ . \hspace{5mm} i=1,2 \ . 
\labbel{F31i2} \ee 
Further, the Wronskian of any two solutions $(i,j)$, 
$$ W^{(i,j)}(x) = F^{(i)}_1(x) F^{(j)}_2(x) -  F^{(i)}_2(x) F^{(j)}_1(x) 
$$ 
is constant (independent of $x$), as the matrix of the coefficients of the 
system is traceless. By using the particular values Eq.(\ref{valKE}) one 
finds 
\be W^{(1,2)}(x) = K(x)K(1-x) - K(x)E(1-x) - E(x) K(1-x) = 
      - \frac{\pi}{2} \ , \labbel{WrKE1-x} \ee 
which is the Legendre relation, and 
\be W^{(1,3)}(x) = \frac{1}{\sqrt{x}}\left[ 
     (1-x)K(x)K\left(\frac{1}{x}\right) + 
    x K(x) E\left(\frac{1}{x}\right) 
    - E(x)K\left(\frac{1}{x}\right) \right] = i \frac{\pi}{2}\,.  \labbel{WrKE/x} \ee 
\newline 

All the integrals $ I^{(a,b)}_i(u), J^{(a,b)}_i(u) $ introduced in 
Appendix \ref{App:AnCont} can be expressed in terms of the complete 
elliptic integrals by using the integral representations of 
(\ref{A5},\ref{A6}) and performing the standard change of the 
integration variable $ b $ into $ t $ according to 
\begin{align}
t^2 = \frac{(b_4-b_2)(b-b_1)}{(b_2-b_1)(b_4-b)}\,,
\end{align}
so that the variable $ x $ entering in the arguments of the resulting 
elliptic integrals reads 
\begin{equation}
x = \frac{(b_2-b_1)(b_4-b_3)}{(b_4-b_2)(b_3-b_1)} \ . 
\end{equation}
More details on the changes of variable in the various regions of $ u $ 
can be found for instance in~\cite{Laporta:2004rb}, and we summarize 
simply the results in the following. 
As a general feature, $ I^{(a,b)}_1(u), J^{(a,b)}_1(u) $ 
are expressed in terms of the first integral $ K(x) $ or $ K(1-x) $, 
while $ I^{(a,b)}_2(u), J^{(a,b)}_2(u) $ involve as well $ E(x), E(1-x) $. 
Note that the term $ \pi(u+3)/3 $, occurring in the definition of 
$ J^{(a,b)}_2(u) $,
see for instance Eq.(\ref{B1}), is compensated by 
a similar term generated by an integration by parts when using the 
above variable $ t $. 

\subsection{The region $0<u<1$}
In this region we have
\begin{align}
I_1^\zu(u) &= \frac{2}{\sqrt{(3-\sqrt{u})(\sqrt{u}+1)^3}}\, K\left( a(u) \right)\,, 
\nonumber \\&\nonumber \\
I_2^\zu(u) &= \frac{15-(u-18)u}{3\, \sqrt{(3-\sqrt{u})(\sqrt{u}+1)^3}} K\left( a(u) \right)
- \sqrt{(3-\sqrt{u})(\sqrt{u}+1)^3} \, E \left( a(u) \right)\,,
\nonumber \\&\nonumber \\
J_1^\zu(u) &= \frac{2}{\sqrt{(3-\sqrt{u})(\sqrt{u}+1)^3}}\, K\left( 1 - a(u) \right)\,, 
\nonumber \\& \nonumber \\
J_2^\zu(u) &= \frac{2(u^2 -12 \sqrt{u} + 3) }{3\, \sqrt{(3-\sqrt{u})(\sqrt{u}+1)^3}} K\left( 1-a(u) \right)
+ \sqrt{(3-\sqrt{u})(\sqrt{u}+1)^3} \, E \left( 1-a(u) \right)\,,
\end{align}
with 
\begin{equation}
a(u) = \frac{16 \sqrt{u}}{(3-\sqrt{u})(\sqrt{u}+1)^3}\ \ , \hspace{1cm} 
1-a(u) = \ \frac{(\sqrt{u}+3)(1-\sqrt{u})^3}{(3-\sqrt{u})(\sqrt{u}+1)^3}\,. 
\labbel{eq:au}
\end{equation}
The Wronskian is 
\begin{align} 
  W^{(0,1)}(u) &= I^{(0,1)}_1(u) J^{(0,1)}_2(u) 
    - I^{(0,1)}_2(u) J^{(0,1)}_1(u) \nonumber\\ 
 &= -2 \Bigl( K(a(u))\,K(1-a(u)) 
   - K(a(u))\,E(1-a(u)) - E(a(u))\,K(1-a(u)) \Bigr) = \pi \ , \nonumber 
\end{align} 
in agreement with Eq.s(\ref{B10},\ref{WrKE1-x}).

\subsection{The region $1<u<9$}
Here we find
\begin{align}
I_1^\un(u) &= \frac{1}{2\,u^{1/4}}\, K\left( b(u) \right)\,, 
\nonumber \\&\nonumber \\
I_2^\un(u) &= \frac{(u^2+12 \sqrt{u} + 3) }{6\, u^{1/4}} K\left( b(u) \right)
- 4\,u^{1/4} \, E \left( b(u) \right)\,,
\nonumber \\&\nonumber \\
J_1^\un(u) &= \frac{1}{2\,u^{1/4}}\, K\left( 1 - b(u) \right)\,, 
\nonumber \\& \nonumber \\
J_2^\un(u) &= \frac{(u^2 -12 \sqrt{u} + 3) }{6\,u^{1/4}} K\left( 1-b(u) \right)
+ 4\,u^{1/4} \, E \left( 1-b(u) \right)\,,
\end{align}
where $b(u)$ is the inverse of $a(u)$, defined in~\eqref{eq:au}, 
\begin{equation}
b(u) = \frac{1}{a(u)}\ = \frac{(3-\sqrt{u})(\sqrt{u}+1)^3}{16 \sqrt{u}}\ \ , 
\hspace{1cm} 
1-b(u) = \frac{(\sqrt{u}+3)(\sqrt{u}-1)^3}{16 \sqrt{u}}\ . \labbel{eq:bu}
\end{equation}
The Wronskian is 
\begin{align} 
  W^{(1,9)}(u) &= I^{(1,9)}_1(u) J^{(1,9)}_2(u) 
    - I^{(1,9)}_2(u) J^{(1,9)}_1(u) \nonumber\\ 
 &= -2 \Bigl( K(b(u))\,K(1-b(u)) 
   - K(b(u))\,E(1-b(u)) - E(b(u))\,K(1-b(u)) \Bigr) = \pi \ , \nonumber 
\end{align} 
in agreement with Eq.s(\ref{B5},\ref{WrKE1-x}).

\subsection{The region $9<u<\infty$}
Above threshold we find
\begin{align}
I_1^\noo(u) &= \frac{2}{\sqrt{(\sqrt{u}+3)(\sqrt{u}-1)^3}}\, K\left( c(u) \right)\,, 
\nonumber \\&\nonumber \\
I_2^\noo(u) &= \frac{2(u^2 + 12\sqrt{u} + 3)}{3\, \sqrt{(\sqrt{u}+3)(\sqrt{u}-1)^3}} K\left( c(u) \right)
- \sqrt{(\sqrt{u}+3)(\sqrt{u}-1)^3} \, E \left( c(u) \right)\,,
\nonumber \\&\nonumber \\
J_1^\noo(u) &= \frac{2}{\sqrt{(\sqrt{u}+3)(\sqrt{u}-1)^3}}\, K\left( 1 - c(u) \right)\,, 
\nonumber \\& \nonumber \\
J_2^\noo(u) &= \frac{15 - (u-18)u }{3\, \sqrt{(\sqrt{u}+3)(\sqrt{u}-1)^3}} K\left( 1-c(u) \right)
+ \sqrt{(\sqrt{u}+3)(\sqrt{u}-1)^3} \, E \left( 1-c(u) \right)\,,
\end{align}
with 
\begin{equation}
c(u) = \frac{(\sqrt{u}-3)(\sqrt{u}+1)^3}{(\sqrt{u}+3)(\sqrt{u}-1)^3}\ \ , 
\hspace{1cm} 
1-c(u) = \frac{16\sqrt{u}}{(\sqrt{u}+3)(\sqrt{u}-1)^3}\ \ . \labbel{eq:cu}
\end{equation}
The Wronskian is 
\begin{align} 
  W^{(9,\infty)}(u) &= I^{(9,\infty)}_1(u) J^{(9,\infty)}_2(u) 
    - I^{(9,\infty)}_2(u) J^{(9,\infty)}_1(u) \nonumber\\ 
 &= -2 \Bigl( K(c(u))\,K(1-c(u)) 
   - K(c(u))\,E(1-c(u)) - E(c(u))\,K(1-c(u)) \Bigr) = \pi \ , \nonumber 
\end{align} 
in agreement with Eq.s(\ref{eq:valWronsk2},\ref{WrKE1-x}). 

\subsection{The region $-\infty < u< 0$}
While in the other three regions, $0<u<1$, $1<u<9$, $9<u<\infty$, the solutions
found in appendix~\ref{App:AnCont} are manifestly real, in this last region, i.e. for euclidean momenta
$u<0$, we have to introduce linear combinations of complex functions in order to get a real result. 
The same thing can be done in terms of elliptic integrals, and one should always recall to give the correct
prescription to $u$, which we assume to be $u + i\,0^+$, also for $u < 0$. With this prescription we obtain

\begin{align}
I_1^\ooz(u) &= 2\,i\,\left[  \frac{K\left( d(u) \right)}{\sqrt{(3-\sqrt{u})(\sqrt{u}+1)^3}} 
-  \frac{K\left( c(u) \right)}{\sqrt{(3+\sqrt{u})(1-\sqrt{u})^3}} \right]\,, 
\nonumber \\&\nonumber \\
I_2^\ooz(u) &=i\, \left\{ \left[ \frac{2(u^2 - 12\sqrt{u} + 3)}{3\, \sqrt{(3-\sqrt{u})(\sqrt{u}+1)^3}} K\left( d(u) \right)
+ \sqrt{(3-\sqrt{u})(\sqrt{u}+1)^3} \, E \left( d(u) \right) \right] \right. \nonumber \\ 
&\quad \; \; - \left.  \left[ \frac{2(u^2 + 12\sqrt{u} + 3)}{3\, \sqrt{(3+\sqrt{u})(1-\sqrt{u})^3}} K\left( c(u) \right)
+ \sqrt{(3+\sqrt{u})(1-\sqrt{u})^3} \, E \left( c(u) \right) \right]
  \right\}\,,
\nonumber \\&\nonumber \\
J_1^\ooz(u) &=   \frac{K\left( d(u) \right)}{\sqrt{(3-\sqrt{u})(\sqrt{u}+1)^3}} 
+  \frac{K\left( c(u) \right)}{\sqrt{(3+\sqrt{u})(1-\sqrt{u})^3}} \,,
\nonumber \\& \nonumber \\
J_2^\ooz(u)  &=\frac{1}{2} \left\{ \left[ \frac{2(u^2 - 12\sqrt{u} + 3)}{3\, \sqrt{(3-\sqrt{u})(\sqrt{u}+1)^3}} K\left( d(u) \right)
+ \sqrt{(3-\sqrt{u})(\sqrt{u}+1)^3} \, E \left( d(u) \right) \right] \right. \nonumber \\ 
&\quad \;\, \; + \left.  \left[ \frac{2(u^2 + 12\sqrt{u} + 3)}{3\, \sqrt{(3+\sqrt{u})(1-\sqrt{u})^3}} K\left( c(u) \right)
+ \sqrt{(3+\sqrt{u})(1-\sqrt{u})^3} \, E \left( c(u) \right) \right]
  \right\}\,,
\end{align}
with $d(u)$ is defined as the inverse of $c(u)$
\begin{equation}
d(u) = \frac{( \sqrt{u}+3)(\sqrt{u}-1)^3}{(\sqrt{u}-3)(\sqrt{u}+1)^3} = \frac{1}{c(u)}\,,
\end{equation}
and $c(u)$ was defined in~\eqref{eq:cu}.
The Wronskian is 
\begin{align} 
  W^{(-\infty,0)}(u) &= I^{(-\infty,0)}_1(u) J^{(-\infty,0)}_2(u) 
    - I^{(-\infty,0)}_2(u) J^{(-\infty,0)}_1(u) \nonumber\\ 
 &= -2\,i\frac{1}{\sqrt{c(u)}} \Bigl( (1-c(u))\,K(c(u))\,K(d(u)) 
   + c(u)\,K(c(u))\,E(d(u)) - E(c(u))\,K(d(u)) \Bigr) = \pi \ , \nonumber 
\end{align} 
in agreement with Eq.s(\ref{eq:valWronsk2},\ref{WrKE/x}). 
On account of Eq.s(\ref{F31i2}), one can express $ K(d(u)), E(d(u)) $ in 
terms of $ K(c(u)), E(c(u)) $ and $ K(1-c(u)), E(1-c(u)) $ , obtaining 
$$ W^{(-\infty,0)}(u) = -2 \Bigl( K(c(u))\,K(1-c(u))
   - K(c(u))\,E(1-c(u)) - E(c(u))\,K(1-c(u)) \Bigr) = \pi \ , $$ 
in agreement with \ref{WrKE1-x}.

\bibliographystyle{bibliostyle}   
\bibliography{Biblio}
\end{document}